\documentclass{aa}  

\usepackage{graphicx}
\usepackage{subcaption}
\usepackage{xcolor}
\usepackage[symbol]{footmisc}
\usepackage{booktabs}
\usepackage{physics}
\usepackage{multirow}
\usepackage{amsmath}
\usepackage{hyperref}
\usepackage{url}
\usepackage{academicons}
\usepackage{svg}
\usepackage{mathrsfs} 
\hypersetup{breaklinks=true,colorlinks=true,urlcolor=blue, linkcolor=blue,  citecolor=blue, bookmarksopen=true}
\DeclareMathSymbol{*}{\mathbin}{symbols}{"03}
\usepackage{orcidlink}
\usepackage{txfonts}

\begin{document}

   \title{Gas dynamics around dust asymmetries in turbulent disks}

   \author{Lizxandra Flores-Rivera 
          \inst{1}\orcidlink{0000-0001-8906-1528}
          \and
          Natascha Manger\orcidlink{0000-0001-6552-8605}
          \inst{2}\fnmsep
          \and
          Michiel Lambrechts 
          \inst{1}\orcidlink{0000-0001-9321-5198}
          \and
          Mario Flock
          \inst{3}\orcidlink{0000-0002-9298-3029}
          \and
          Sebastian Lorek
          \inst{2}\orcidlink{0000-0002-5572-4036}
          \and
          Anders Johansen
          \inst{2}\orcidlink{0000-0002-5893-6165}
          \and
          Hubert Klahr
          \inst{3}\orcidlink{0000-0002-8227-5467}
          }

   \institute{Centre for Star and Planet Formation, Globe Institute, University of  Copenhagen, \\
    \O ster Voldgade 5-7,   
    1350 Copenhagen, Denmark
              \email{lizxandra.rivera@sund.ku.dk}
         \and
             Center for Computational Astrophysics, Flatiron Institute, 162 Fifth Ave, New York, NY 10010, USA
        \and
            Max Planck Institut f\"ur Astronomie, K\"onigstuhl 17, 69117 Heidelberg, Germany\\
             }

   \date{---}

\abstract
  {High-resolution ALMA observations have revealed asymmetric dust crescents in several protoplanetary disks, suggesting efficient dust trapping mechanisms potentially linked to gas vortices. While such features have been associated with vortices—whether induced by massive planets, turbulence, or other disk processes—their origin remains unclear. In this study, we investigate the viability of dust trapping by vortices that are self-sustained in disks dominated by Vertical Shear Instability (VSI) turbulence. We perform 3D hydrodynamic simulations using the \textsc{PLUTO} code with Lagrangian particles of three sizes (1 mm, 500~$\mu$m, 100~$\mu$m) to analyze the gas–dust dynamics around vortices. Our simulations reveal the formation of multiple vortices, including two characteristic large-scale, long-lived vortices that are able to capture the dust particles. We also find that dust vertical diffusion is reduced within vortices, suggesting that these structures preferentially enhance radial and azimuthal motions. Finally we generate synthetic dust continuum images at different wavelength bands and velocity residuals to compare the observable properties with ALMA observations. No clear spiral features are observed in either the synthetic dust images or the velocity residuals, unlike in vortices triggered by planets. Projection effects at high disk inclinations can obscure dust asymmetries, implying that more disks may host dust crescents than currently reported. }


   \maketitle
 
\section{Introduction}

In recent years, high-resolution observations of protoplanetary disks by the Atacama Large Millimeter/Submillimeter Array (ALMA) have revealed striking dust substructures, in the form of rings, gaps, spirals, and asymmetric crescent-shaped features \citep{Bae_2023}. Observations in the sub-millimeter (sub-mm) dust continuum of dust crescents suggest that large dust particles are being trapped in the disk midplane \citep{van_der_Marel_2013, Perez_2018}. Determining which mechanism is primarily responsible for the observed dust asymmetries remains a significant challenge. 

The most widely accepted physical mechanism for generating gas vortices is the Rossby Wave Instability \citep[RWI;][]{Lovelace_1999, Li_2000, Li_2001}. This instability is triggered by a local minimum in the radial profile of the potential vorticity (or vortensity), defined as $\eta = \frac{\kappa^{2}S^{-2/\gamma}}{2\Omega_{k}\Sigma_\mathrm{gas}}$, where $\kappa$ is the epicyclic frequency, which describes the oscillation frequency of a fluid element undergoing small radial displacements about a circular orbit and is given by $\kappa = R^{-3}\frac{\partial(R^{4}\Omega_{k}^{2})}{\partial R}$; $\Omega_{k}$ is the Keplerian angular velocity; $\Sigma_\mathrm{gas}$ is the gas surface density; $S = P/\Sigma_\mathrm{gas}^{\gamma}$ denotes the entropy with $P$ the gas pressure; and $\gamma$ is the adiabatic index. This quantity depends on the disk’s density, temperature, and rotational structure, governed by the balance of radial forces including the centrifugal, Coriolis, and pressure gradient forces. The RWI leads to the formation of anti-cyclonic vortices at planetary gap edges, provided the planet is sufficiently massive—comparable to a Jupiter- or Neptune-mass planet \citep{val-borro_2007, Lyra_2009a}. Vortices may also arise near disk boundaries where sharp pressure or viscosity transitions generate long-lived pressure maxima, often observed as dusty rings in ALMA observations \citep{Lin_2014}. Additionally, RWI-induced vortices can form in disks perturbed by a binary companion on an eccentric orbit \citep[i.e.,][]{Price_2018}.

The RWI can also emerge in turbulent disk environments driven by the Vertical Shear Instability \citep[VSI;][]{Nelson_2013}. The VSI is a hydrodynamical (HD) instability in accretion disks that arises from vertical gradients in rotational velocity, leading to velocity perturbations and turbulence in both the vertical and radial directions \citep{Barker_2015, Stoll_2016, Lyra_2019}. \citet{Richard_2016} first demonstrated that small vortices can form in disks susceptible to the VSI. Later, \citet{Latter_2018} confirmed that the Kelvin-Helmholtz (KH) instability, acting as a parasitic instability in VSI modes, are responsible of the small scale vortices. \citet{Richard_2016} also suggested that secondary instabilities—such as the RWI or the elliptical instability \citep[E.I.;][]{Lesur_2009} may ultimately determine the fate of these vortices, an idea later summarized by \citet{Lyra_2019}. Building on subsequent vortex studies in VSI scenarios, \citet{Manger_2018} showed that the VSI can generate large‑scale, long‑lived vortices—azimuthally elongated and vertically aligned—as a secondary instability triggered by the RWI \citep[see also][]{Manger_2020, Pfeil_2021}. While such vortices have been commonly reported in VSI simulations under locally isothermal, recent high‑resolution simulations by \citet{Shariff_2024}, with up to 70 cells per scale height, were the first to show that large-scale vortices are likely absent due to the enhanced dynamical range of kinetic energy cascading to smaller scales. This resolution effect was later confirmed by \citet{Lesur_2025}, who also reported the absence of long‑lived midplane vortices and attributed this to the action of the EI. These results highlight ongoing uncertainties regarding how resolution and gas‑cooling timescales influence vortex formation in VSI‑driven turbulence. 

It remains unclear which mechanism initiates the observed dust asymmetries—whether the RWI acts as a secondary instability in VSI-dominated disks, or whether a massive planet first carves a gap that subsequently triggers a vortex responsible for the dust crescent.
Hydrodynamical simulations by \citet{Huang_2019} have shown that spiral arms generated by RWI-driven vortices exhibit surface density contrasts comparable to those produced by sub-thermal mass planets but are too weak to be detected in scattered light, suggesting that the prominent near-infrared spirals observed in systems such as SAO 206462 and MWC 758 are more likely planet-induced.
To distinguish between these mechanisms and their observational manifestations, it is essential to analyze the associated gas dynamics.

Vortices have been widely investigated as efficient dust traps in protoplanetary disks, where local pressure maxima can concentrate solids and potentially trigger planetesimal formation \citep{Johansen_2004, Lyra_2009a}. RWI vortices have been studied in the context of planet–disk interaction simulations that include Lagrangian dust particles, confirming their ability to trap dust until gravitational collapse leads to the formation of bodies with masses up to that of a super-Earth \citep{Lyra_2009a, Fu_2014, Lin_2014, Bae_2016, Ma_2025}. Vortices formed at the edges of planet-induced gaps can become significantly elongated in azimuth when the planet grows over sufficiently long timescales, a feature that may help distinguish them observationally from other asymmetry-generating mechanisms \citep{Hammer_2017, Hammer_2019}. The lifetime of such vortices can extend to several thousand orbits when dust coagulation and fragmentation are taken into account \citep{Fu_2014}. While previous studies have explored VSI vortices in gas-only simulations \citep{Richard_2016, Manger_2018, Manger_2021, Pfeil_2021}, only few works \citep{Flock_2020, Blanco_2021, Huang_2025} have conducted 3D simulations demonstrating that VSI-driven turbulence can concentrate dust into RWI vortices and small-scale clumps that is related to the KH instability.

From an observational standpoint, the recent exoALMA Large Program \citep{Teague_2025} has presented new high-resolution gas observations including four of the disks that exhibit prominent dust crescents in the sub-mm continuum. These observations achieve a high angular resolution of 100 mas (approximately 14 au at the typical distances of the sources) and a spectral resolution of 26 m~s$^{-1}$, enabling the detection of subtle structures and motions that offer critical insights into dust asymmetries and their role as physical processes governing planet formation. In \citet{Wolfer_2025}, the authors analyzed the line-of-sight velocities of $^{12}$CO $(J=3-2)$ and $^{13}$CO $(J=3-2)$ and identified deviations from purely Keplerian rotation around the four disks with dust crescents. Vortex diagnostics in simulations reveal distinct azimuthally asymmetric red-blue velocity patterns used to probe deviations from Keplerian motion \citep[e.g.,][]{Robert_2020}. In their work, vortices formed at the edge of gas-depleted regions produced apparent spiral-arm features in the gas line-of-sight velocity maps. However, the interpretation of such features in observations remains challenging due to the complexity of gas dynamics, where multiple processes (e.g., other disk instabilities or binary interactions) may obscure or mimic vortex signatures.

\begin{table*}\centering
\caption{Model parameters}  \label{tb:model_parameters}
\scalebox{0.9}{
\begin{tabular}{lcccccccr}
\toprule
Model & $r_\mathrm{in,out}/R_0$ & $\theta$ & $\phi$ & Grid size$(N_r \times N_\theta \times N_\phi)$ & $p$ & $q$ & $H/R$ & $\beta_\mathrm{cool}\Omega_{k}^{-1}$  \\ 
\midrule 
q-0.5tau1e-4 & 0.5-2.0 & 1.2208-1.9208 & 0.0-6.2831853 & $512\times256\times1024$ & -2.25 & -0.5 & 0.1 & $10^{-4}$ \\
\bottomrule
\end{tabular}}
\end{table*}

\begin{figure}[htp!]
\centering
\includegraphics[width=7cm]{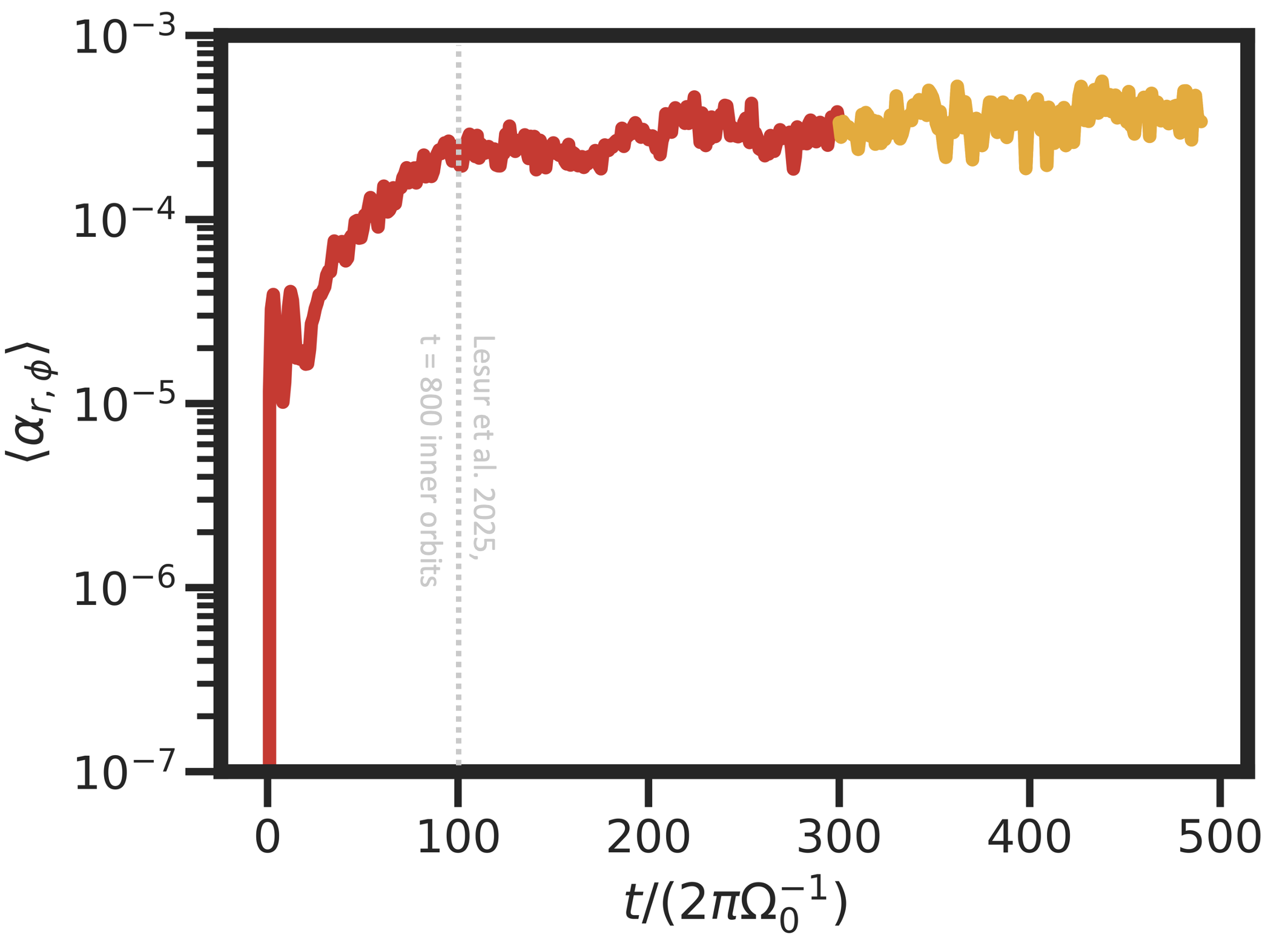}
\includegraphics[width=9cm]{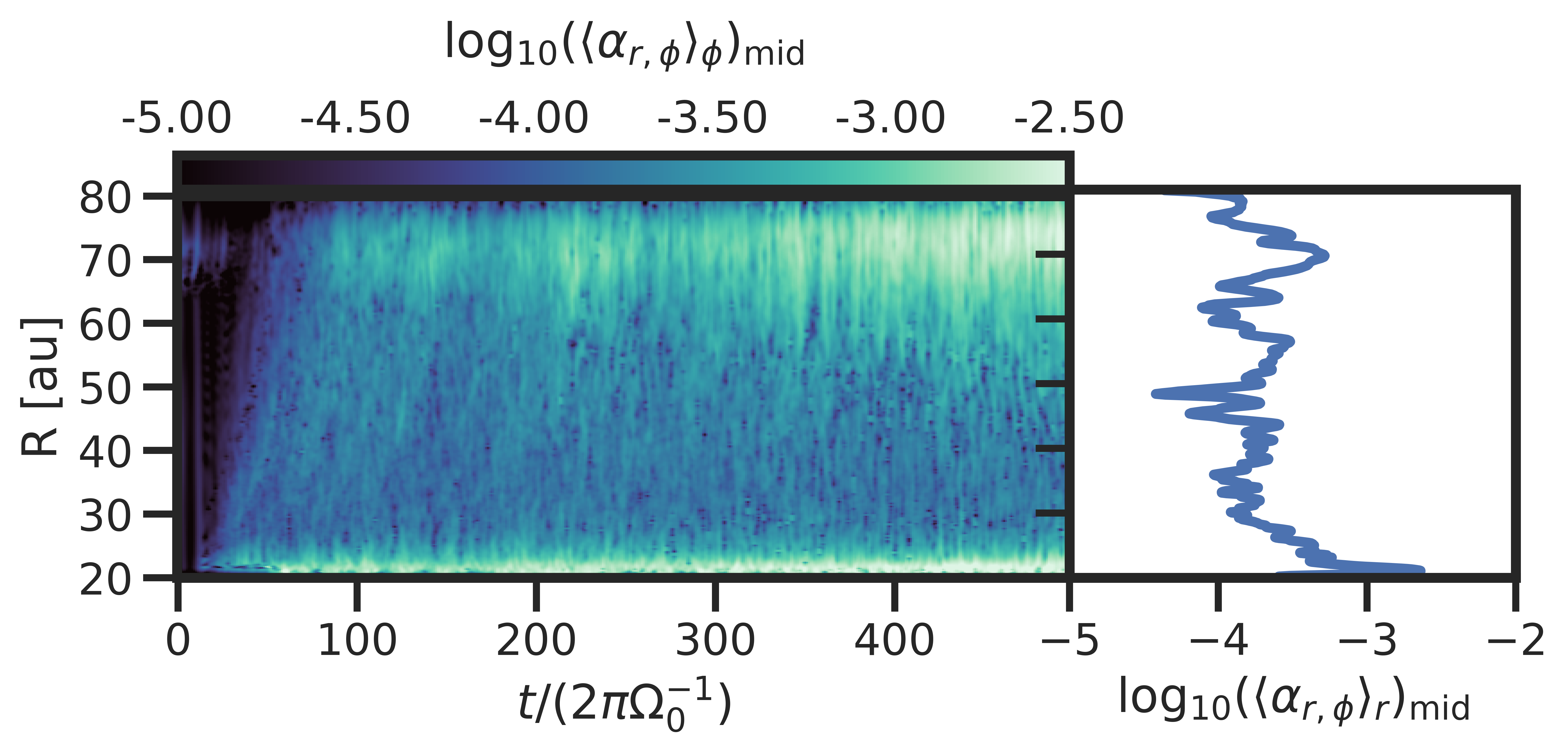}
\caption{Reynolds stress-to-pressure ratio, $\alpha_{r,\phi}$. The top panel shows the time-averaged $\alpha_{r,\phi}$, where the yellow line represents the convergence value, $\langle \alpha_{r,\phi} \rangle = 3\times10^{-4}$. For reference, the vertical dotted line corresponds to the outer local orbit at 0.25$R_{0}$ used by \citet{Lesur_2025} in their analysis, as shown in their Figure 1. The bottom-left panel displays the time series of $\alpha_{r,\phi}$ in r-direction, averaged over azimuth in the 3D VSI-active disk. The bottom-right panel presents the radial Reynolds stress-to-pressure ratio over the time series and averaged in the azimuthal directions.}
\label{fig:alpha_all}
\end{figure} 

In this paper, we investigate the influence of VSI-driven, self-sustained vortices on dust distribution, gas kinematics, and their potential observational signatures in protoplanetary disks. In Section \ref{sec:gas}, we describe the three-dimensional simulation setup for both gas and dust. Section \ref{sec:results} presents the results, focusing on the locations within the disk where particles become concentrated and trapped inside the vortices (see Appendix \ref{sec:criterion} for the RWI vortex criterion). We use a single time snapshot to test whether large-scale dust concentrations at similar locations motivated by those observed in the MWC 758 disk \citep{Dong_2018}, can arise from vortices triggered by the RWI in a turbulent disk driven by the VSI. However, we are not specifically concerned with matching the models solely to the properties of MWC 758. In general, we aim to link observed sub-mm asymmetries to vortex-induced dust trapping from our models. We then demonstrate how post-processing the simulation data allows us to diagnose dust asymmetries around vortices, analyze gas kinematic perturbations from Keplerian motion, and predict how these features would appear in ALMA observations. In Section \ref{sec:discussions}, we evaluate the feasibility of explaining observed dust asymmetries in disks based on our model predictions. Finally, our conclusions are summarized in Section \ref{sec:conclusions}.

\section{Methods}
\label{sec:gas}

We performed 3D simulations of the gas in the protoplanetary disk based on the principles of mass, momentum, and energy conservation using the PLUTO\footnote{\textsc{PLUTO 4.3} is an open source code available for download: \url{http://plutocode.ph.unito.it/download.html}} code \citep{Mignone_2007}. The simulations are carried in spherical coordinates $(r,\theta,\phi)$ using a uniform spaced grid in $\theta$ and $\phi$, and logarithmic spaced grid in $r$. The fluxes in each cell are computed using the parabolic reconstruction method using fourth order with the Harten, Lax, Van Leer (HLLC) solver. For the time integration we use the third-order Runge-Kutta algorithm with the Courant-Friedrichs-Lewy (CFL) number set to 0.3. The grid extends from 0.5$R_0$ to 2$R_0$ in the radial direction, where $R_0 = 1$ in code units, and from $\pi/2 - 0.35$ to $\pi/2 + 0.35$ (i.e., $\pm 3.5H$) in the meridional direction. The analysis are presented in cylindrical coordinates $(R, \phi, Z)$ and are scaled to 40 au in physical units. This numerical setup was chosen to achieve a resolution of approximately 37 cells per scale height in all directions (Table \ref{tb:model_parameters}). Our disk configuration follows the equilibrium solutions of the density and rotational velocity profile in cylindrical coordinates based on \citet{Nelson_2013}

\begin{equation}
    \label{density}
      \rho(R,Z) = \rho_{\mathrm{0}} \left(\frac{R}{R_{0}} \right)^{p}  \exp \left (\frac{-Z^{2}}{2H^{2}} \right)  \,, 
\end{equation}

\begin{equation}
      \label{azimuthalvelocity}
      \Omega(R,Z) = \Omega_\mathrm{k} \left[ (p+q) \left(\frac{H}{R} \right)^{2} + (1+q) - \frac{qR}{\sqrt{R^2 + Z^2} } \right]^{1/2}  ,
\end{equation}

\noindent The variables $p$ and $q$ corresponds to the exponents in the radial profiles of temperature and midplane density, respectively. The initial midplane density is given by $\rho_\mathrm{0} = \frac{\Sigma_{\mathrm{0}}}{\sqrt{2\pi}~H/R R_{0}}$, where $\Sigma_{\mathrm{0}}=30$ g~cm$^{-2}$ is the initial surface density at the reference radius $R_\mathrm{0}$, and $H/R$ is the aspect ratio (see Table \ref{tb:model_parameters}). The scale height profile of the gas is $H \propto R^{(q+3)/2}$. 

The gas is modeled as an ideal gas, with internal energy $e$ related to temperature and density via the caloric equation of state. The pressure is given by $P = \rho e/(\gamma - 1)$, where $\gamma = 1.4$ is the adiabatic index appropriate for diatomic molecular hydrogen. The gas pressure is defined as $P=c_{\mathrm{s}}^{2}\rho$, where $c_{\mathrm{s}} \propto (\frac{R}{R_\mathrm{0}})^{q}$ is the radially varying isothermal sound speed. The temperature profile is related to the gas pressure based on $c_{\mathrm{s}}$ as $T\propto c_\mathrm{s}^{2}$ and the sound speed is related to the gas pressure scale height as $c_\mathrm{s} = H \Omega_\mathrm{k}$.

\begin{figure}[htp!]
\centering
\includegraphics[width=9cm]{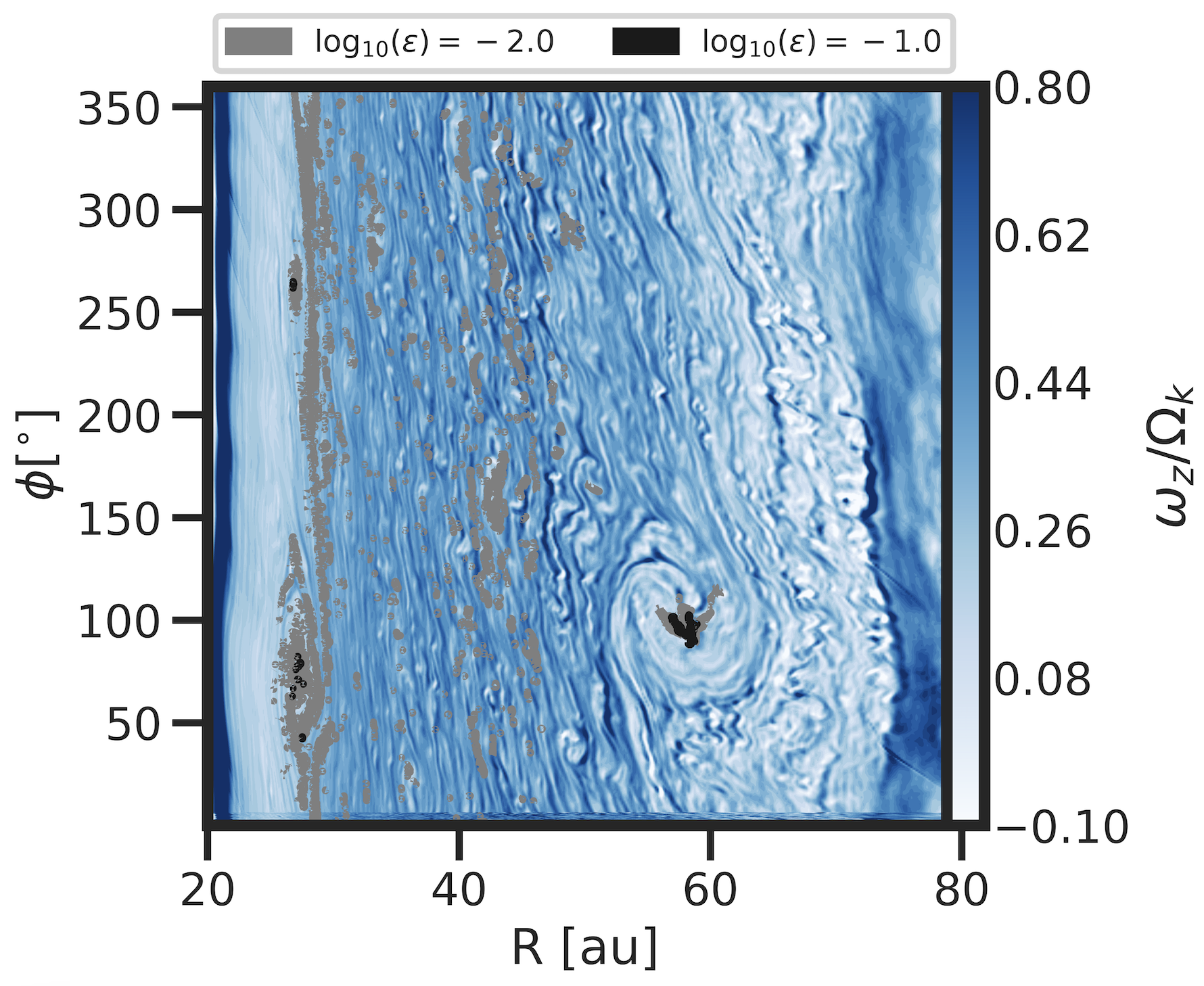}
\caption{Vorticity of the gas in the midplane in units of the keplerian frequency at 350 orbits. Overplotted are two dust-to-gas mass ratio values of 1 mm particles in the midplane.}
\label{fig:vorticity}
\end{figure}

\begin{figure*}[htp!]
\centering
\includegraphics[width=15.2cm]{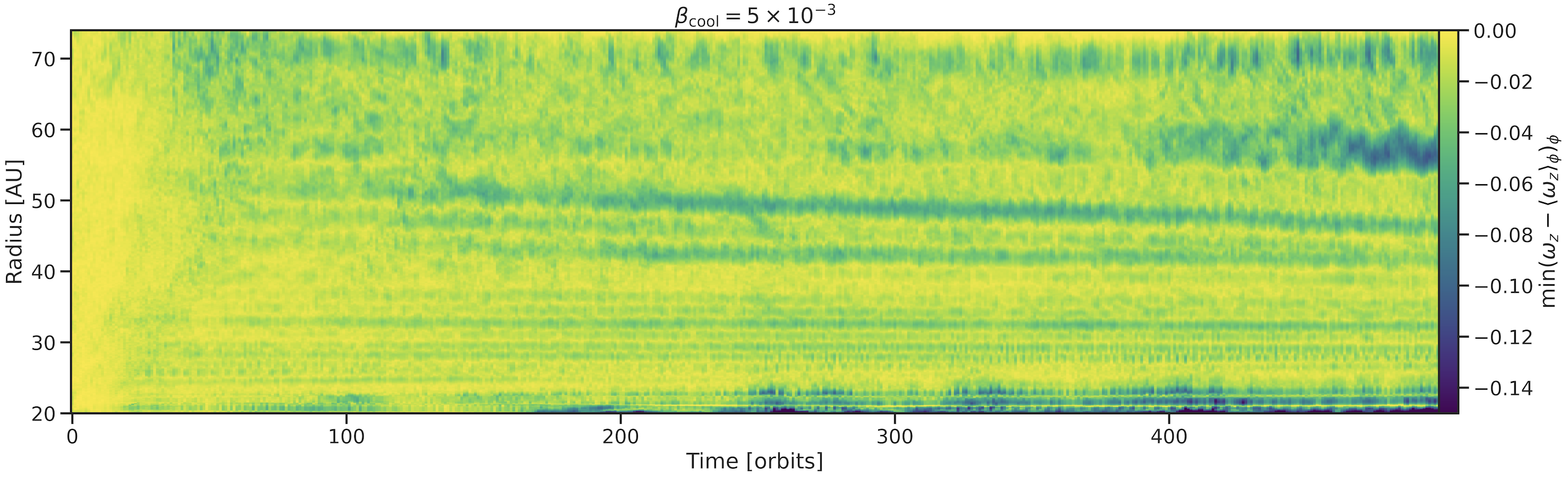}
\caption{Time evolution of the minimum vorticity residual as a function of radius, illustrating the radial position and migration of vortices in the VSI-unstable disk. The horizontal dashed lines outline the final locations of the two large scale vortices.}
\label{fig:vort_residual}
\end{figure*}

We use a simple thermal relaxation or cooling relaxation scheme in which the gas pressure evolves towards an equilibrium state over a characteristic cooling timescale, $t_\mathrm{cool} = \beta_\mathrm{cool} \Omega_{k}^{-1}$. The thermal relaxation is represented by the time evolution of the pressure dependent of the local density and the local sound speed

\begin{equation}
\label{eq:cooling}
\frac{dP}{dt}= -\frac{P - \rho c_\mathrm{\textit{s},ini}^{2}}{\beta_\mathrm{cool}\Omega_{k}^{-1}},
\end{equation}
 
\noindent which is relaxed to its initial value described by $c_\mathrm{\textit{s},ini}$. The initial disk conditions adopt values typical of T Tauri systems, with $q=-0.5$, which has been shown to favor VSI growth \citep{Manger_2021}, and $p=-2.25$, a choice that ensures a constant mass accretion rate is maintained \citep[see equation 13 in][]{Manger_2021}, consistent both with theoretical expectations \citep{Lynden-Bell_1974} and with observational constraints \citep[e.g.,][]{Wolf_2003, Sauter_2009}. We adopt a cooling time of $10^{-4}\Omega_{k}^{-1}$, which is sufficiently short to permit the development of the VSI in the outer disk (Table \ref{tb:model_parameters}). 

We adopt the same boundary conditions as in \citet{Manger_2018} to avoid wave reflections and minimize boundary effects at the inner and outer edges of the domain. Specifically, within buffer zones of width $\Delta R = 1.0 H$ and $\Delta \theta = 0.05$, the density and velocity components are damped toward their initial values when the velocity normal to the boundary points inward into the domain. We use periodic boundaries in the azimuthal direction. Within these buffer zones, the damping timescale is set relative to the local orbital period as

\begin{equation}
\label{eq:damp}
\frac{dv_{x}}{dt}= -\frac{v_{x} - v_{x,0}}{\tau_\mathrm{damp}} \cdot f^{2}.
\end{equation}

\noindent Here, $v_{x}$ represents any of the primitive variables, such as density or velocity components. The damping timescale is defined as $\tau_\mathrm{damp} = 0.1 \frac{2\pi}{\Omega_\mathrm{k}}$, and the damping factor is given by $f = \frac{R - R_\mathrm{boun}}{\Delta R}$ in the radial direction, and by $f = \max\left(\frac{R - R_\mathrm{boun}}{\Delta R}, \frac{\theta - \theta_\mathrm{boun}}{\Delta \theta}\right)$ in the meridional direction. The parameters $R_\mathrm{boun}$ and $\theta_\mathrm{boun}$ denote the radial and meridional locations of the boundaries of the damping layer within the buffer zone. 

\section{Dust}
\label{sec:init_cond}

In this paper, we are interested in understanding where dust-loading  concentration occurs in VSI-active protoplanetary disks and determining qualitatively whether it has observational implications. From the 2D ($R,Z$) model of the density and rotational velocity described in Section \ref{sec:gas},  we introduced three populations of Lagrangian particles with sizes of 100 $\mu$m, 500 $\mu$m, and 1 mm in the simulation. We reset the dust component midway through the simulation runtime. This was necessary because, by approximately 200 orbits, the initial set of dust particles had already drifted inward before fully developed vortices—whose full development occurs beyond 250 orbits—could effectively trap incoming particles. This approach allowed us to investigate and analyze the role of vortices in dust trapping more thoroughly. The particle dynamics is described using a Lagrangian dust module \citep{Mignone_2019} in the PLUTO code using a spherical coordinates system. These dust grains are assumed to be homogeneous, compact, and spherical with a dust density of, $\rho_\mathrm{dust} = $1.7 g~cm$^{-3}$. We do not include the effects of the dust-to-gas feedback \citep[see, e.g., studies of dust-to-gas feedback in VSI simulations by][]{Schafer_2020, Schafer_2022}. We assume that dust particles are sufficiently small compared to the gas mean free path, allowing treatment within the Epstein regime \citep{Epstein_1924}. In this regime, the Stokes number is given by $\mathrm{St} = t_s \Omega_\mathrm{k}$, where $t_s$ is the stopping time—the timescale over which a dust grain adjusts to gas motion, depending on its size. It dictates the time that it takes for a dust particle to be slowed down by the gas flow and in terms of the disk properties is defined as

\begin{figure*}[htp!]
\centering
\includegraphics[width=15.cm]{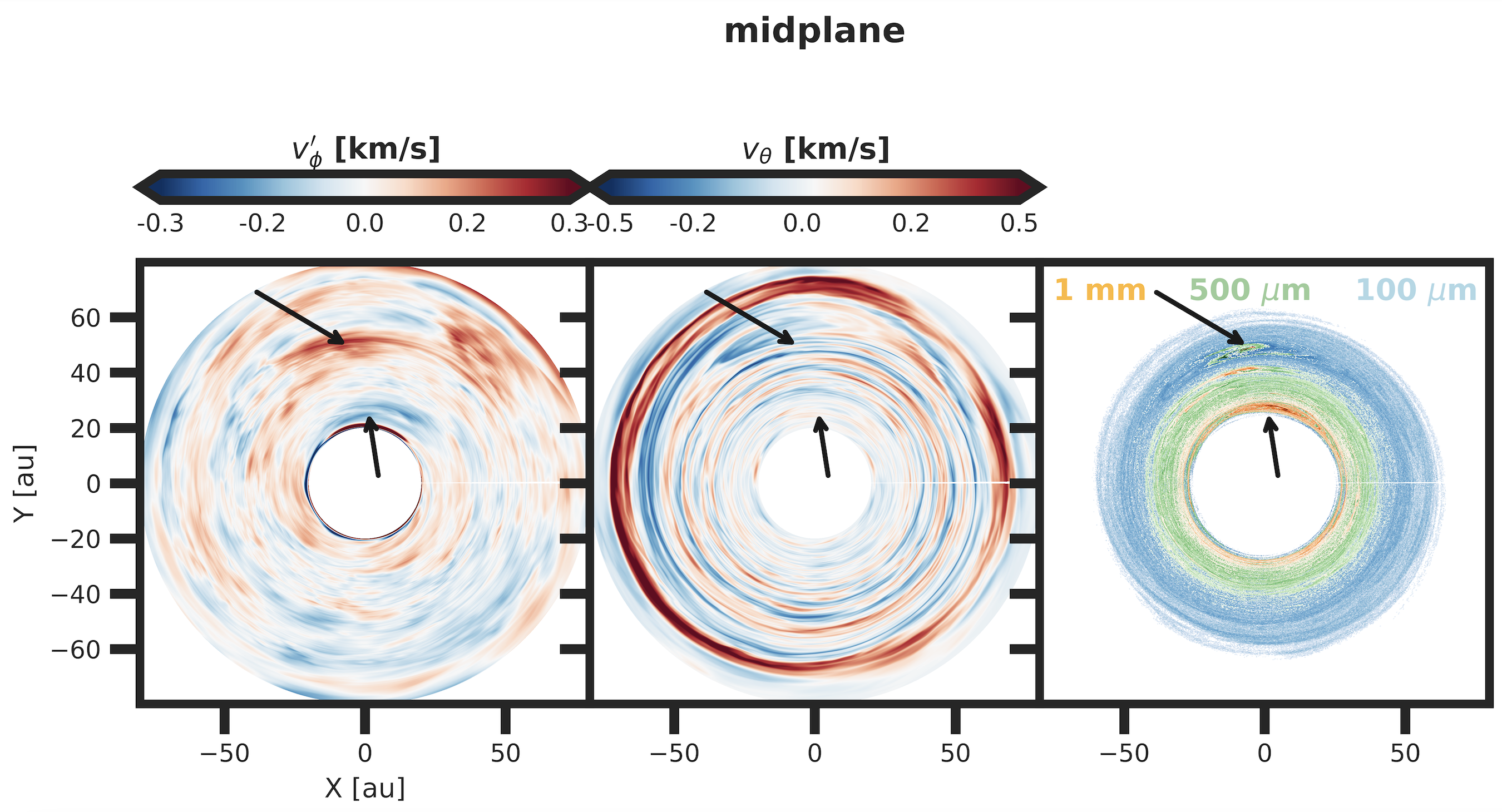}
\includegraphics[width=15.cm]{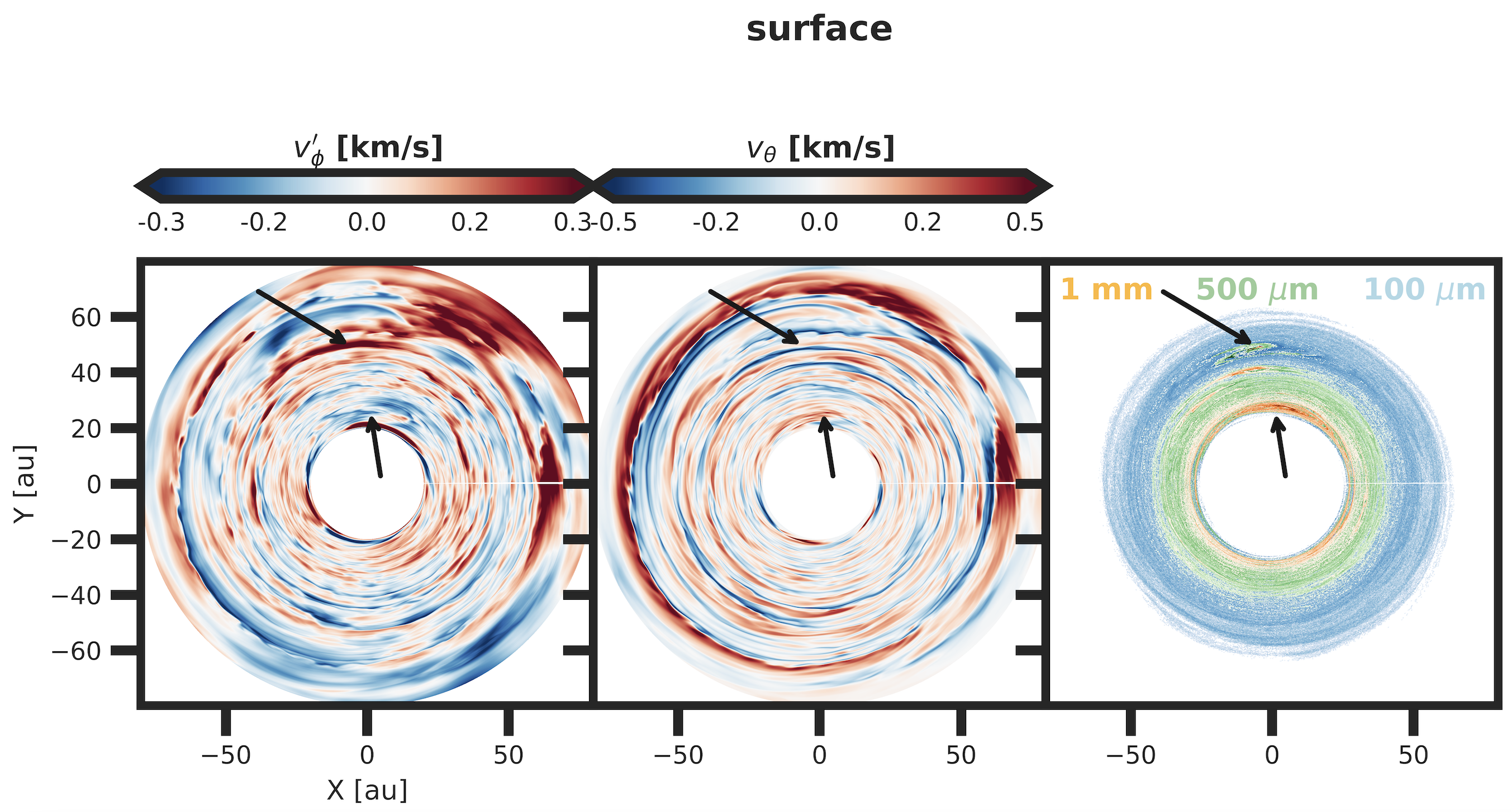}
\caption{Pertubational azimuthal velocity (top and bottom left panels), vertical velocity (top and bottom middle panels) of the disk at the midplane ($z=0$, top panels), and close to the disk surface ($z>2H$, bottom panels). To make it easier to identify the two large-scale vortices in the disk, we show the dust distribution vertically averaged in the top and bottom right panels. The black arrows point at the location of the inner and outer vortices.}
\label{fig:velocities+dust}
\end{figure*}

\begin{equation}
    \label{stopping_time}
      t_{s} = \frac{\mathrm{St}}{\Omega_\mathrm{k}} = \frac{a\rho_\mathrm{dust}}{\rho(R,Z) H \Omega_\mathrm{k}} = \frac{a\rho_\mathrm{dust}}{\rho(R,Z) c_{s}}\, , 
\end{equation}

\noindent where $a$ is the dust particle size, which is kept constant throughout the dynamical evolution of the disk (see Appendix \ref{sec:stokes} for the Stokes number values at different scale heights in the disk for each grain size). Aside from the Lagrangian particle sizes, which are considered large in our grain sample, we include small grains between 0.1 $\mu$m and 30 $\mu$m discretized in 10 bins important to consider when post-processing the simulation data in the RADCM3D code (see Section \ref{sec:syn_im}). These small grains are not modeled as Lagrangian particles in the VSI simulations, as they remain well coupled to the gas. For the dust mass distribution we follow the approach as in \citet{Ruge_2016} where the dust-size distribution is defined based on the MRN \citep{Mathis_1977} distribution with a power-law of -3.5. The MRN distribution fits well for the larger particles that dominate the intensity at the chosen wavelengths (see Section \ref{sec:syn_im}). The small grains are very well-coupled with the gas and spatially distributed uniformly. We insert a million of particles and for the mentioned distribution we obtain dust-to-gas mass ratio values for the simulation run: $M_\mathrm{small,d}/M_\mathrm{gas}=0.0003$, $M_\mathrm{100 \mu m,d}/M_\mathrm{gas}=0.0012$, $M_\mathrm{500 \mu m,d}/M_\mathrm{gas}=0.0026$, and $M_\mathrm{1mm,d}/M_\mathrm{gas}=0.0036$.

\section{Results}
\label{sec:results}

To asses the time-integrated turbulence in the disk, Figure \ref{fig:alpha_all} (top) shows the time evolution of volume-weighted Reynolds stress-to-pressure ratio, $\alpha_{r,\phi}$, that is radially and azimuthally averaged 

\begin{equation}
    \label{eq:alpha}
      \alpha_{r,\phi} = \frac{\int_{}^{}T_{r,\phi}dV}{\int{}^{}PdV} = \frac{\int_{}^{}\rho v_{r}^\prime v_{\phi}^\prime dV}{\int{}^{} \rho c_s^{2} dV}\, .
\end{equation}

\noindent Here $v_{r}^\prime$ and $v_{\phi}^\prime$ are the velocity perturbations calculated by subtracting the
azimuthally averaged value at each $r$ and $\theta$. The saturation phase occurs around 100 orbits, during which $\alpha_{r,\phi}$ converges to a value of $3 \times 10^{-4}$, and the turbulence remains stable until approximately 500 orbits. We use a single snapshot at 350 orbits that shows the two large-scale dust crescents at similar locations motivated by those observed in the MWC 758 disk \citep{Dong_2018}. Figure \ref{fig:alpha_all} (bottom) shows the time series of $\alpha_{r,\phi}$ in the midplane, where an increasing trend is observed over time.

\subsection{Vorticity and dust concentration}

To identify dynamical gas structures acting as dust traps, we generate midplane vorticity maps in the $R-\phi$ plane. The vertical component of the vorticity, normalized by the Keplerian frequency, is given by

\begin{equation} 
\label{eq:vorticity}
\frac{\omega_{z}}{\Omega_k} = (\nabla \times \Vec{v})_{z}
           = \frac{1}{\Omega_k} \left( 
            \frac{1}{r} \frac{\partial r v_\theta}{\partial r} 
            - \frac{1}{r} \frac{\partial v_r}{\partial \phi}
           \right) ,
\end{equation} 

\begin{figure*}[htp!]
\centering
\includegraphics[width=9cm]{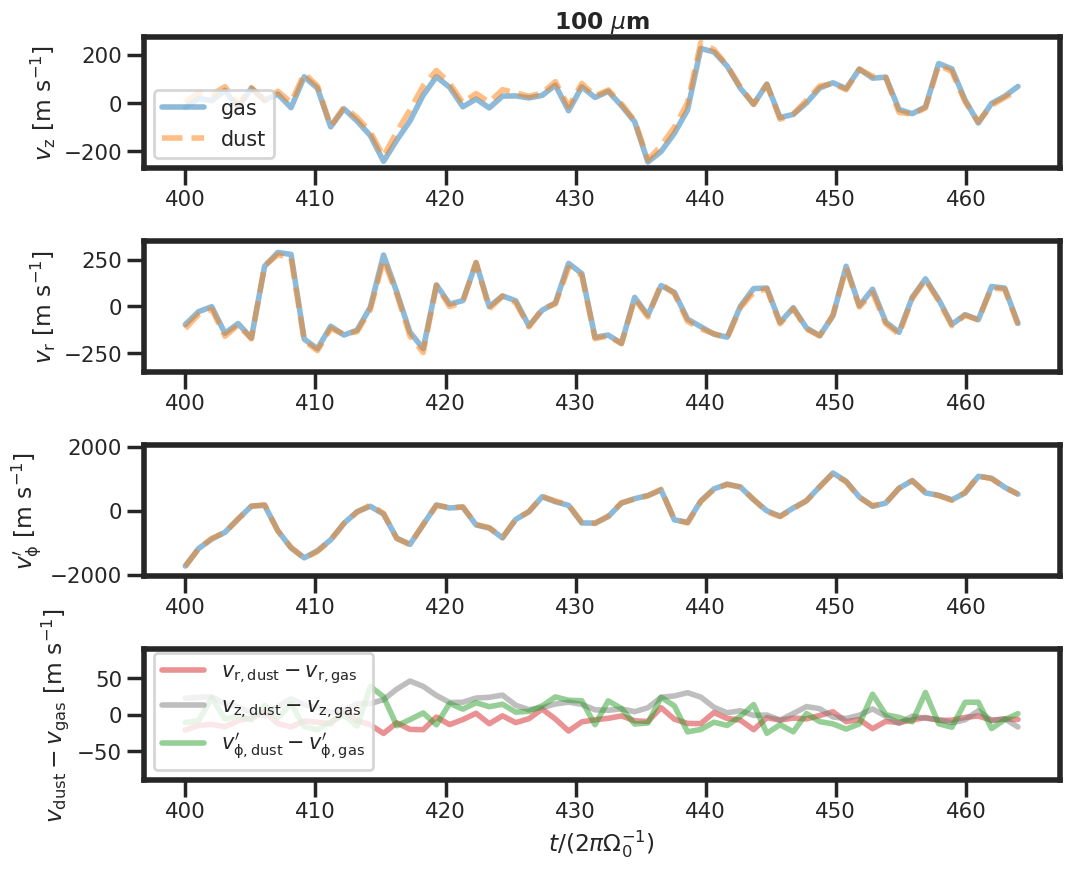}
\includegraphics[width=9cm]{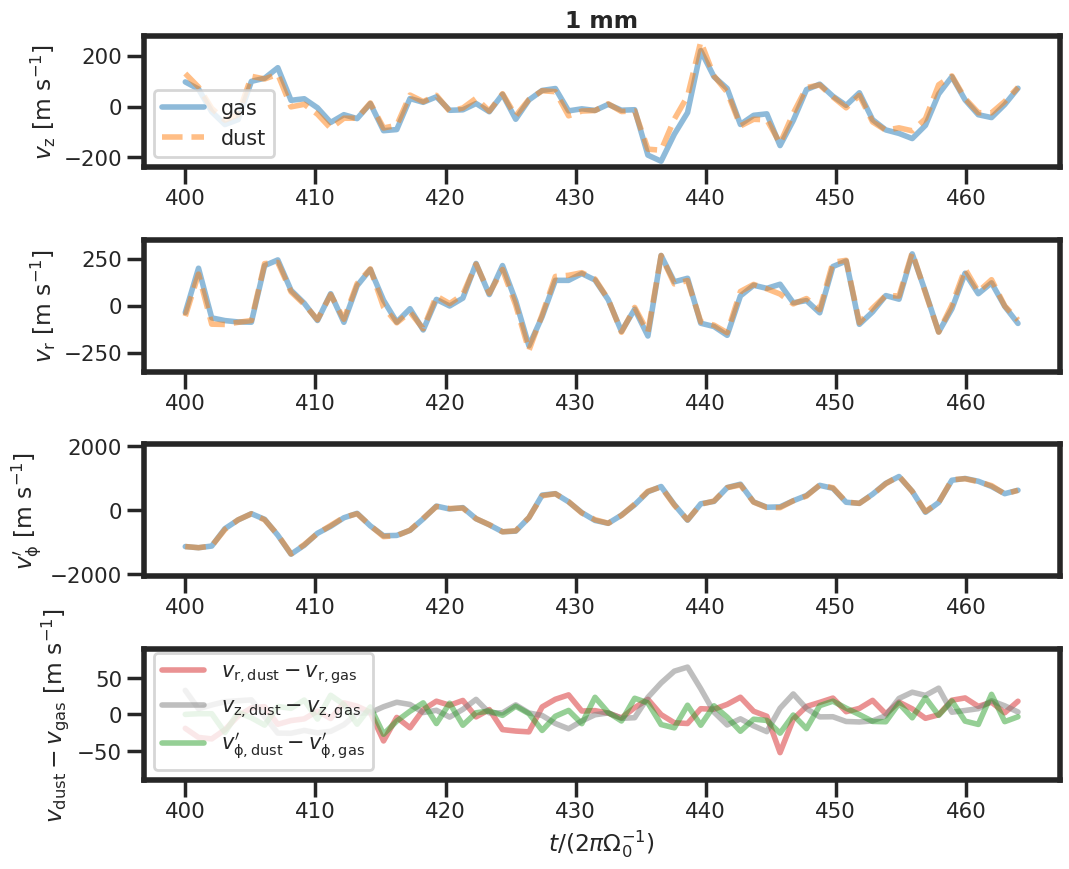}
\caption{Velocity profile of a single 100~$\mu$m-sized particle (left) and a 1 mm-sized particle (right) within the vortex.}
\label{fig:veloc_profile}
\end{figure*} 

\noindent where $\Vec{v}$ is the gas velocity field \citep{Lovelace_1999}. Figure \ref{fig:vorticity} shows the resulting midplane vorticity map normalized by $\Omega_k$, overlaid with contours of the dust-to-gas ratio ($\mathrm{log}_{10}\varepsilon=-1.0$ and $-2.0$) for 1 mm particles at the midplane simply to illustrate where these larger particles tend to concentrate. Two large-scale anti-cyclonic vortices are evident at distinct azimuthal locations and they exhibit strong particle accumulation toward their centers.

To demonstrate the evolution of vortices that are self-sustained within the VSI-unstable disk, we generated a time-resolved vorticity residual map (Figure \ref{fig:vort_residual}). This was constructed by applying a box filter to the midplane vertical vorticity, $\omega_z$, using a filter width of 200 grid cells in the azimuthal direction and 30 cells radially (approximately two pressure scale heights). The filtering suppresses small-scale fluctuations, enhancing coherent vortex structures. We then computed the residual as the deviation of the smoothed local vorticity from its azimuthal average within each annulus, allowing us to isolate vortex signatures from large-scale azimuthal flows.

The inner vortex forms near a local pressure maximum (see Appendix \ref{sec:more_properties} for more information about the pressure perturbations) at the inner damping zone (see Section \ref{sec:gas}), where VSI turbulence is intentionally suppressed to produce a low-viscosity transition region. Our vorticity residual analysis (Figure \ref{fig:vort_residual}) shows that the vortex persists away from the damping region, indicating that it is most likely a self-sustained structure within the VSI flow. The outer vortex becomes fully developed within 60 au after approximately 250 orbits and gradually migrates inward at a rate of $\sim$0.04 au per orbit ($\approx 9\times10^{-4} R_{0}$ per orbit) because of the steep radial pressure gradient \citep[see more in][]{Manger_2018}. We notice a vorticity field perturbation appearing around $\sim$100 orbits that seems to be associated with the origin of the outer vortex’s trailing structures, but no obvious vortex signatures appear in the snapshots between 50 and 100 orbits. Because its formation occurs after 250 orbits, when most 1 mm dust particles have already drifted inward, it captures fewer particles than the inner vortex. Its aspect ratio, $\chi \gtrsim 7$, is roughly half that of the inner vortex. Additional small-scale vortex trails are also visible in Figure \ref{fig:vort_residual}. These smaller structures can transiently concentrate particles—particularly those with sizes of 500 $\mu$m and 1 mm—but we do not investigate them further in this study, as our analysis focuses on the largest, long-lived vortices that dominate the disk dynamics.

We additionally verified that vortices are self-sustained in VSI disks using a model with the same density and temperature gradients but a different cooling parameter, $\beta_{\mathrm{cool}} \Omega_{k}^{-1} = 5\times10^{-3}$ (see Appendix \ref{sec:vort_residual_model3}). The vorticity residual evolution in this model also exhibits coherent vortex signatures forming away from the damping zone, confirming that vortex generation is not restricted to the fiducial boundary setup.

\subsection{Azimuthal and vertical velocity fields}

To examine how the vortices modify the local gas dynamics, we analyzed the azimuthal and vertical velocity perturbations. Figure \ref{fig:velocities+dust} shows the azimuthal and vertical gas velocity perturbations together with the vertically averaged dust distribution for the same snapshot as in Figure \ref{fig:vorticity}. The inner and outer vortices are traced by concentrations of 1 mm and 100 $\mu$m particles, respectively (indicated by arrows). Both vortices produce strong azimuthal velocity deviations from Keplerian rotation, exhibiting characteristic red–blue dipolar patterns that indicate anti-cyclonic motion. The outer vortex, located near 50 au, shows the clearest retrograde flow signature, with sub-Keplerian motion on one side and super-Keplerian motion on the other. These perturbations persist from the midplane to the disk surface ($>2H$), though they become more pronounced at higher altitudes (Figure \ref{fig:velocities+dust}, bottom panel). Vertical velocities remain weak near the vortices but increase toward the surface, where the flow becomes more asymmetric and turbulent. The presence of velocity perturbations caused by vortices significantly alters the surrounding gas kinematics.

\subsection{Dynamical trajectory of particles inside a vortex}

\begin{figure*} [htp!]
\includegraphics[width=18cm]{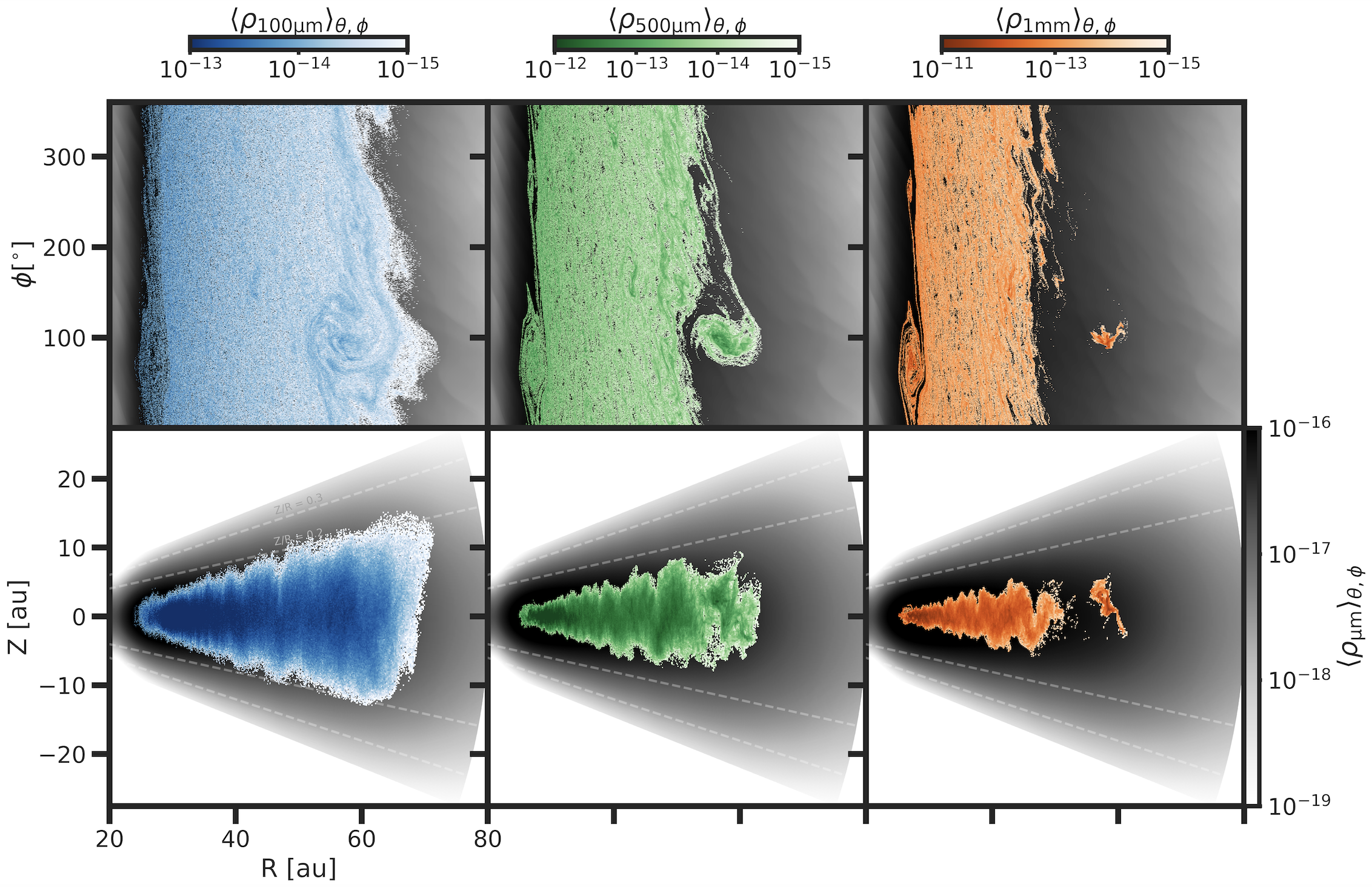} 
\caption{Dust density distribution. Top panels. Dust density distribution vertically averaged at 350 orbits in the $\phi$-R plane. The large grains are 100 $\mu$m (blue), 500 $\mu$m (green), and  1 mm (orange). The small grains spanning from 0.1-30 $\mu$m is plotted as a black shade in the disk and are assumed to be well-coupled with the gas. Bottom panels. Same dust density distribution as top panels but viewed in the Z-R plane and averaged in the azimuthal direction.}
\label{fig:dust_dens_azimuth}
\end{figure*}

\begin{figure}[htp!]
\centering
\includegraphics[width=8cm]{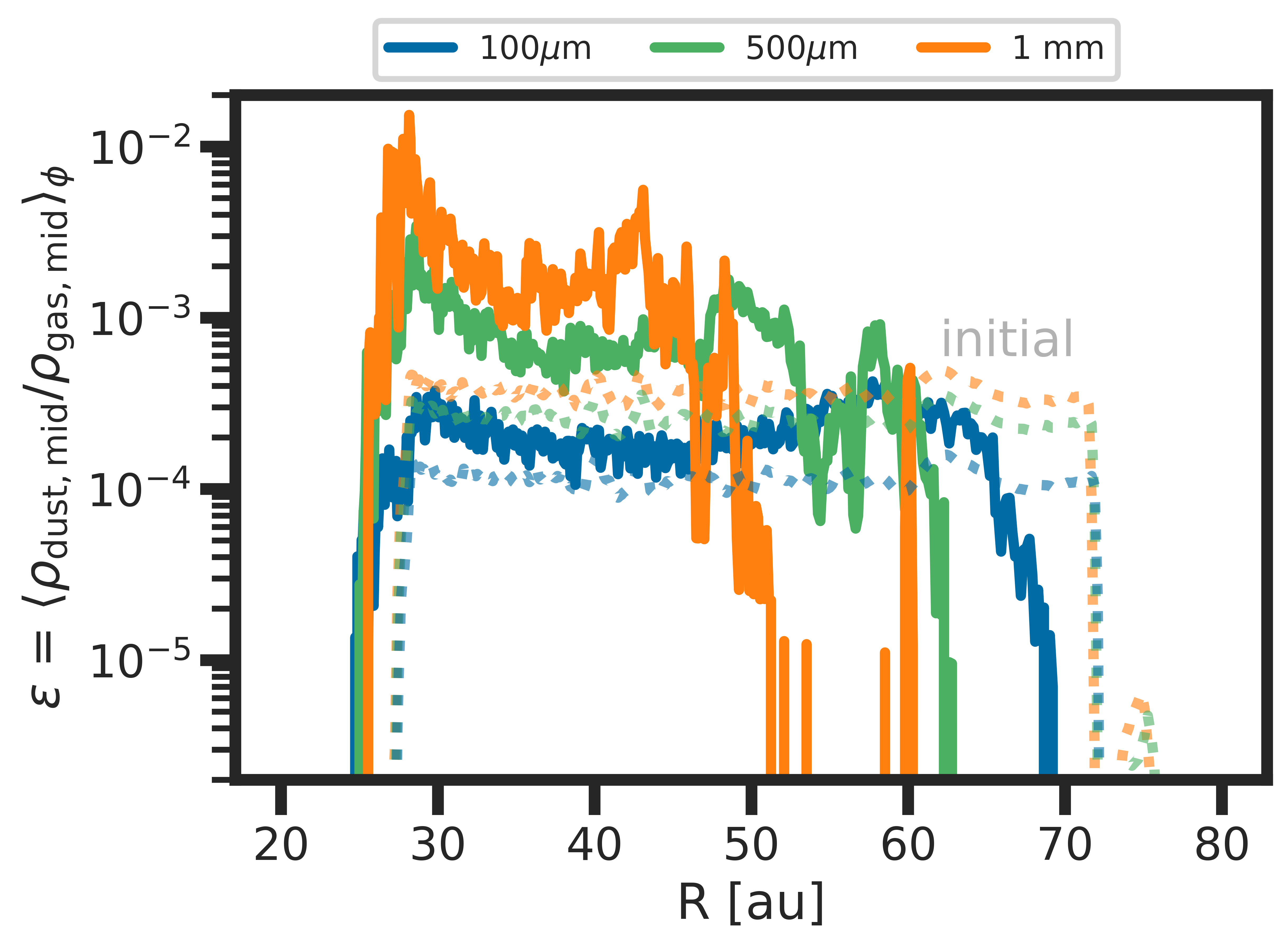}
\caption{Dust-to-gas mass ratio at the midplane at 350 orbits. The dotted lines represent the initial dust-to-gas mass ratio value for each grain size.}
\label{fig:e_all}
\end{figure} 

We further analyzed the trajectory of randomly selected particles, representative of a 100~$\mu$m-sized and 1~mm sized grain, to assess its velocity profile after being captured by the outer vortex (see Appendix \ref{sec:3D_trajectory} for their spatial 3D trajectories). To reduce computational expense, the analysis focuses on a limited orbital interval between 400 and 460 orbits, once the particles are already confined within the vortex. In Figure~\ref{fig:veloc_profile}, we present the particle’s velocity components over time for a 100~$\mu$m-sized grain (left) and 1~mm sized grain (right). In both particles' profile, we find that the vertical velocity no longer exhibits the characteristic upward and downward symmetry typically expected in VSI-dominated regions, as shown in Figure 6 and Figure 7 in \citet{Flores-Rivera_2025}, whereas the radial and perturbed azimuthal velocities largely retain their sinusoidal behavior. Additionally, we observe that the vertical velocity becomes subdominant, with a smaller amplitude compared to the radial component. 

We also conducted a test analysis of the trajectories of a 100 $\mu$m- and a 1 mm-sized particles trapped within the inner vortex. In both particles' cases, the vertical velocity amplitudes are nearly a factor of two lower than those observed for particles in the outer vortex. Nevertheless, we observe a consistent trend in which the sinusoidal symmetry of the vertical velocity is broken and remains subdominant relative to the radial component.

\subsection{Spatial distribution of the dust}
\label{sect:dust_dist}

\begin{figure}[htp!]
\centering
\includegraphics[width=8cm]{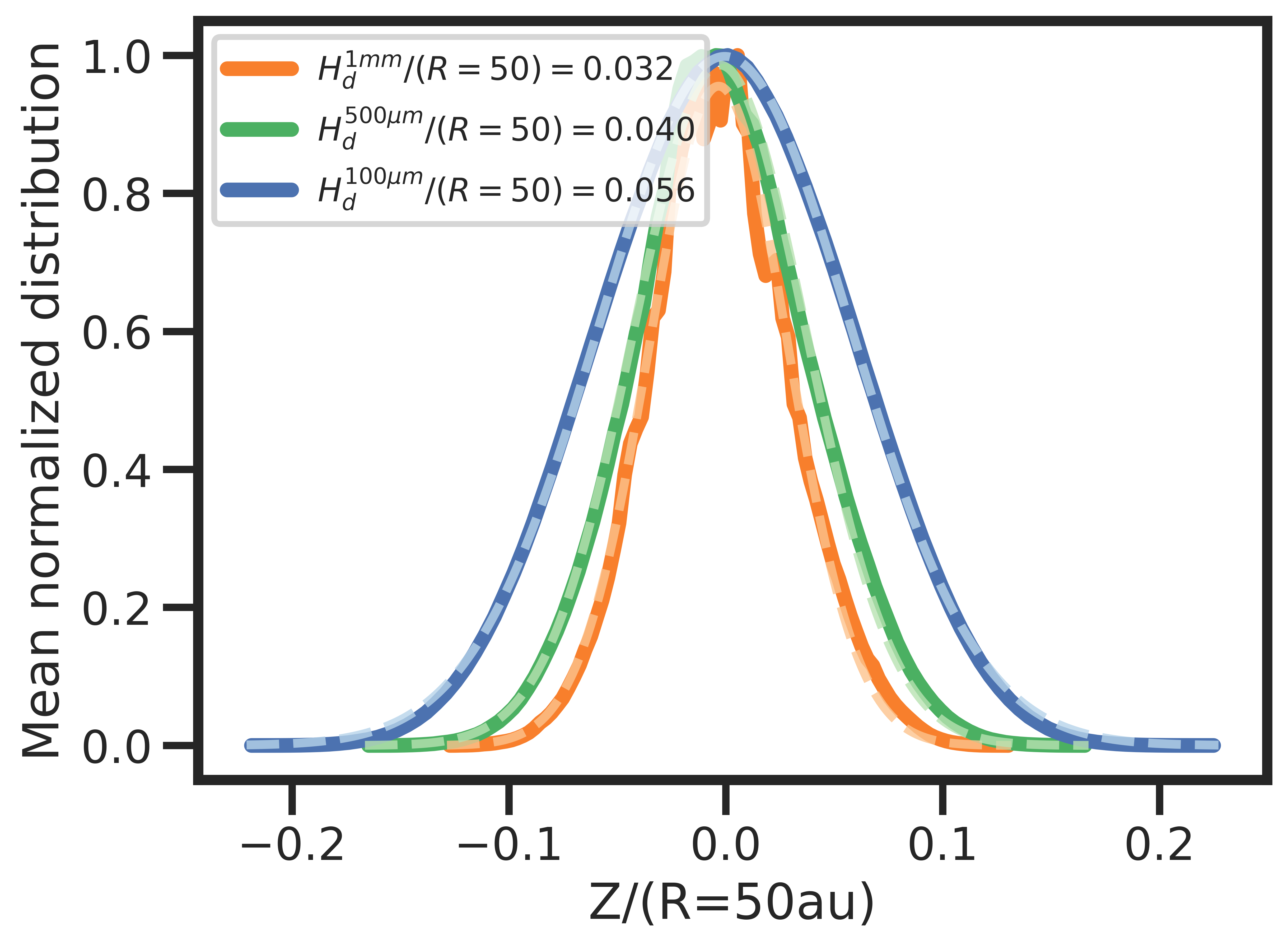}
\caption{Mean vertical dust distribution, $Z$, averaged over the radial range 40–50 au and over time after 300 orbits. Colored lines show the mean normalized distributions for 1 mm (orange), 500 $\mu$m (green), and 100 $\mu$m (blue) grain sizes. The standard deviation of each distribution corresponds to the dust scale height ($H_d$) for the respective grain size.}
\label{fig:dust_gauss}
\end{figure}

To examine the spatial distribution of solids, Figure \ref{fig:dust_dens_azimuth} shows the dust density for three grain sizes in the $\phi$–$R$ and $Z$–$R$ planes at the same snapshot as in Figure \ref{fig:vorticity}. The $100~\mu$m particles remain broadly distributed and well coupled to the gas, whereas the 1 mm grains occupy smaller disk radii and are concentrated within the inner vortex. The intermediate $500~\mu$m particles are trapped in both the inner and outer vortices, where the later captures primarily the $100~\mu$m and $500~\mu$m grains after most 1 mm particles have drifted inward. Notably, small-scale vortices outside the pressure maxima also contribute to localized dust concentrations, especially for 1 mm particles (Figure \ref{fig:dust_dens_azimuth}, right panel).

The vortices are more difficult to visualize in the vertical distribution of particles (Figure \ref{fig:dust_dens_azimuth}, bottom row). The vertical dust distribution of the 1 mm particles shows that these particles are more settled toward the midplane ($Z/R<0.1$), while $500~\mu$m grains exhibit slightly greater vertical extent. At the outer vortex location, the $500~\mu$m particles form slightly more compact radial and vertical concentration, whereas the $100~\mu$m grains appear more dispersed. In contrast, the inner vortex is less prominent in the vertical dust profile, with the 1 mm distribution appearing relatively flat. Figure \ref{fig:e_all} confirms these trends, showing azimuthally averaged midplane dust-to-gas mass ratios: the 1 mm and $500~\mu$m particles exhibit the highest enhancements, consistent with the vortex trapping locations.

To quantify the vertical distribution of solids, we measured the vertical position, $Z$, of all particles radially averaged between 40 and 50 au, after 300 orbits. Figure \ref{fig:dust_gauss} shows the normalized vertical dust distributions for each grain size (solid lines), with Gaussian fits overplotted (dashed lines). At 50 au, we find $H_{d}^{1\mathrm{mm}}/R=0.0325\pm0.0003$, $H_{d}^{500\mu \mathrm{m}}/R=0.0404\pm0.0002$, and $H_{d}^{100\mu \mathrm{m}}/R=0.0562\pm0.0001$, yielding dust-to-gas scale height ratios of $H_{d}/H = 0.32$, 0.40, and 0.56, respectively.

\begin{figure*}[htp!]
\centering
\includegraphics[width=15cm]{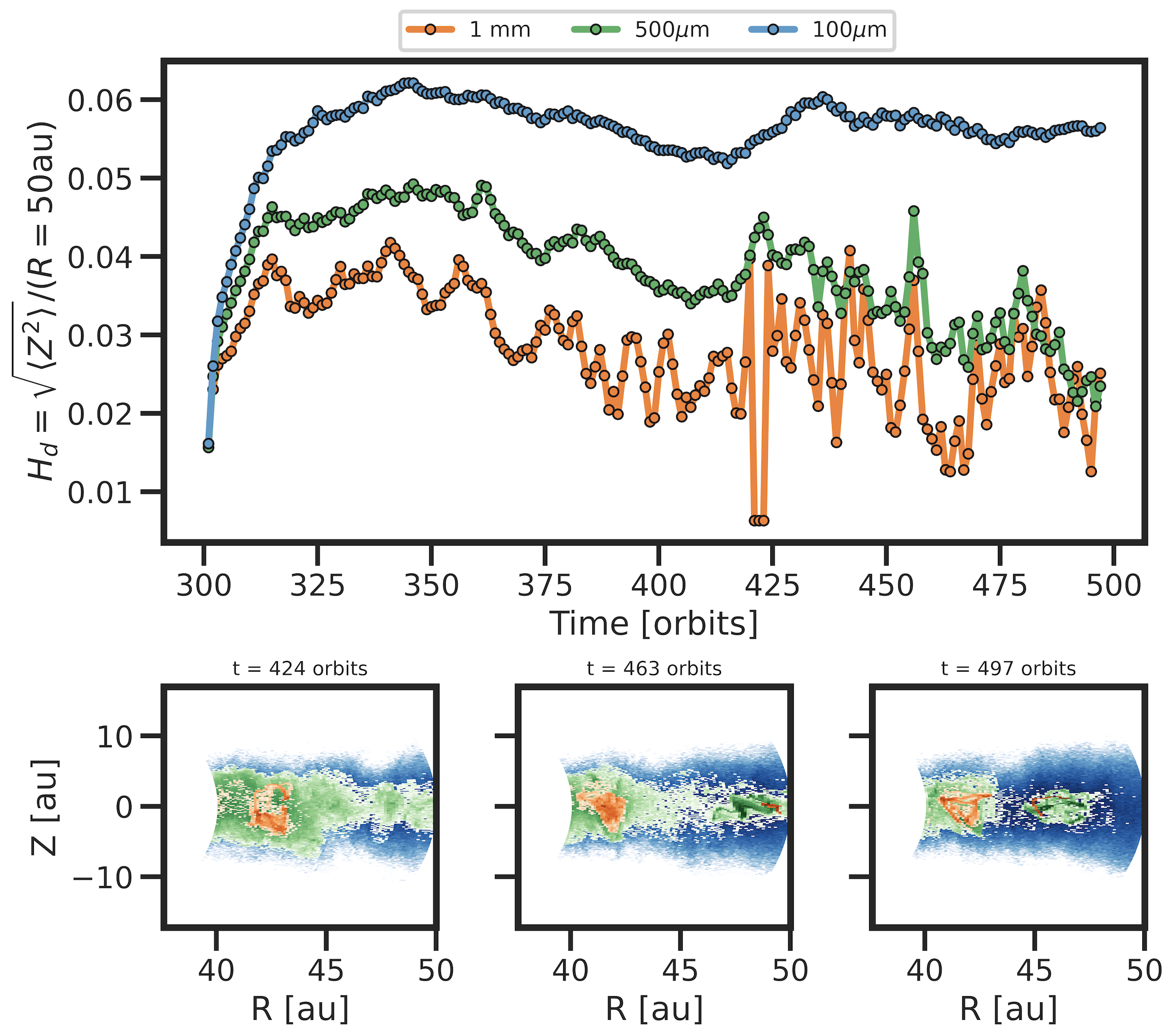}
\caption{Dust scale height over time for all particles species. The dust scale height is radially averaged from 40 au to 50 au. Each marker represents the averaged dust scale height calculated by taking the rms of the vertical position of the particles for all sizes per orbit. Strong vertical fluctuations are observed for the 1 mm and 500 $\mu$m grain sizes after approximately 424 orbits. This corresponds to a regime where the dust distribution is characterized by localized and compact concentrations within vortices (see the orange region in the bottom three panels at three different snapshots).}
\label{fig:dust_scale_height_over_time}
\end{figure*}

The time evolution of the dust scale height, shown in Figure \ref{fig:dust_scale_height_over_time}, reveals that $500~\mu$m and 1 mm particles gradually settle and drift inward, while the $100~\mu$m grains remain more vertically extended. After $\sim$350 orbits, both larger grain sizes exhibit strong fluctuations in $H_d$, with additional variations around $\sim$420 orbits. We note that the strong fluctuations seen after $\sim$450 orbits are not caused by an artificial inverse cascade or a breakdown of the turbulence. Instead, they arise when the large outer vortex migrates into the 40–50 au measurement region, temporarily dominating the dust distribution and producing non‑axisymmetric vertical oscillations. These oscillations arise from vertically stirred, compact dust concentrations within vortices, as seen in the $R$–$Z$ and $R$–$\phi$ projections (bottom panels of Figure \ref{fig:dust_scale_height_over_time}).

During these fluctuation phases, the vertical dust distribution deviates from a Gaussian profile, either forming a sharply peaked layer or a bimodal structure. In such cases, Gaussian fits fail to capture the true vertical extent, making the rms of the particle vertical positions a more robust diagnostic of $H_d$. After $\sim$460 orbits, the 1 mm particles have largely drifted inward, leaving only small remnants trapped in the vortices, visible as narrow, non-Gaussian dust layers (orange regions in Figure \ref{fig:dust_scale_height_over_time}).

\section{Synthetic images at different wavelength bands}
\label{sec:syn_im}

In order to generate the intensity maps of the dust continuum emission of our disk model, we use \textsc{RADMC-3D}\footnote{\url{https://www.ita.uni-heidelberg.de/~dullemond/software/radmc-3d/manual_radmc3d/installation.html}} \citep{Dullemond_2012}. The stellar parameters used are characteristic of a T Tauri star to explore the general physical conditions under which such features may arise. We use $T_{\star}$ = 4000 K, $M_{\star}$ = 0.5 M$_{\odot}$, and $R_{\star}$ = 2 R$_{\odot}$.

We generated the dustkapscatmat$\_$dsharp$\_*$.inp files that contains Mie-based scattering matrices for compact, spherical grains made of 36$\%$ water, 17$\%$ astronomical silicates, 3$\%$ troilite, and 44$\%$ refractory organics by volume \citep{Birnstiel_2018}. The opacity values are made by the \textsc{DSHARP$\_$OPAC} package using the Mie code of Bohren and Huffman \citep[BHMIE;][]{Bohren_1983} embedded in a python script. The opacities are computed for all the dust populations considered here, ranging from 0.1 $\mu$m to 1 mm, under the standard power-law distribution. We adopted anisotropic scattering by setting scattering$\_$mode = 2 in the radmc3d.inp file, enabling the Henyey–Greenstein approximation using the scattering opacities and their corresponding anisotropy parameters provided in the dust opacity file.   

\begin{figure*}
\centering
\includegraphics[width=17.2cm]{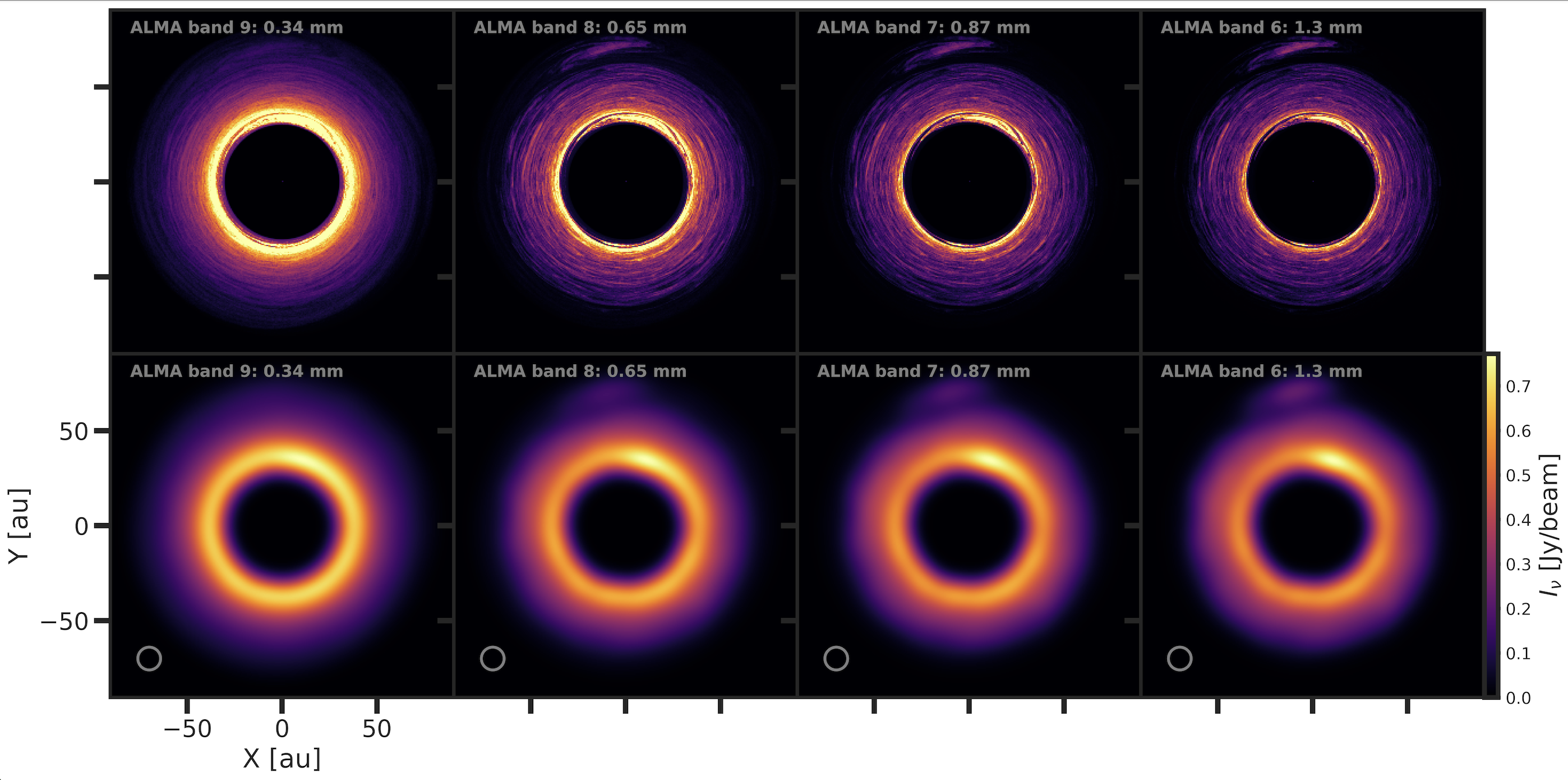}
\caption{Synthetic images of the simulated disk at 0$^{\circ}$ inclination. Top row shows the brightness of the simulated disk after radiative transfer for four different ALMA wavelength bands (columns). Bottom row shows the convolution of the disk images using a circular beam with the FWHM of 12 au ($\sigma_\mathrm{beam}$ = 5 au). Overall, the presence of two asymmetric dust features are clearly seen at the inner ring and in the outer parts of the dust continuum emission. The inner one is found to be at the same location as the dust concentration.}
\label{fig:radtrans_0incl}
\end{figure*}

\begin{figure*}[htp!]
\centering
\includegraphics[width=17.2cm]{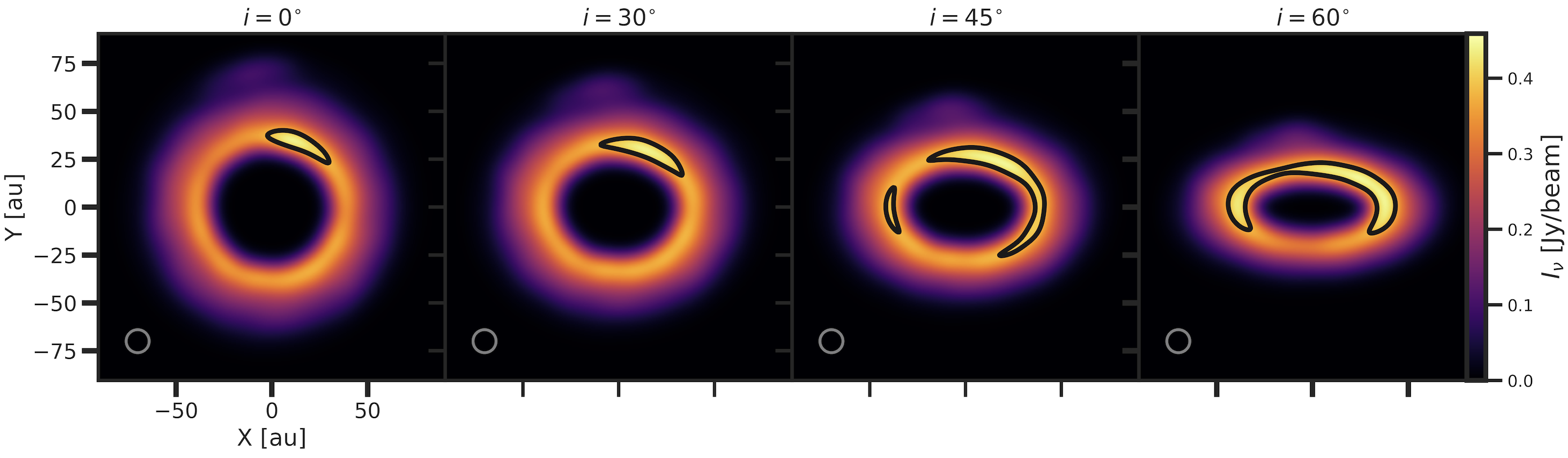}
\caption{Dust crescent at different disk inclinations. The black contour represents 85$\%$ of the peak intensity for the ALMA band 6 frequency. The elongation of the dust crescent increases with disk inclination due to an imaging projection effect. This effect is also evident in the other ALMA wavelength bands.}
\label{fig:effect_incl}
\end{figure*}

The synthetic images of our modeled protoplanetary disk, are shown in Figure \ref{fig:radtrans_0incl} (top panels) at four different ALMA wavelength bands: 340 $\mu$m (Band 9), 650 $\mu$m (Band 8), 870 $\mu$m (Band 7), and 1.3 mm (Band 6). As a test, we also generated synthetic images at longer wavelengths, including 3.0 mm (the lowest-frequency end of the Next Generation Very Large Array, ngVLA), 7.0 mm, and 10 mm (the highest-frequency end of the ngVLA). At these wavelengths, the two bright crescent-like features remain prominent, but the emission appears more compact, consistent with dust trapping in the inner and outer vortices identified in previous figures. The rest of the disk becomes fainter, as the ngVLA bands are most sensitive to grains between 5 and 15 mm in size—particles not included as Lagrangian components in our simulations—making it noteworthy that the modeled 1 mm grains still produce sufficient flux to weakly reveal the vortices. The dust crescents are most prominent in Bands 8, 7, and 6. At shorter wavelengths, such as Band 9, the brightness contrast of the crescents is considerably reduced compared to the longer-wavelength bands. Moreover, the outer crescent is no longer visible, and spiral arms are also absent, consistent with the lack of spirals previously seen in the dust density distribution. 

The bottom panels of Figure \ref{fig:radtrans_0incl} show the synthetic intensity maps at the same wavelengths bands but convolved with a synthetic 2D circular Gaussian beam of $\sigma_\mathrm{beam}$ = 5 au (equal to a circular beam with FWHM = 12 au, similar to the synthesized beam used in \citet{Dong_2018} for MWC 758). The previously sharp and intricate substructures observed in the unconvolved images such as multiple narrow rings, crescents, and azimuthal asymmetries, are now smoothed out due to the finite beam size. Despite the convolution, prominent features remain visible in Bands 8, 7, and 6, where the two asymmetric crescents are still present. In Band 9, the inner dust crescent appears over-brightened but takes on a more ring-like morphology due to the lower brightness contrast at this wavelength, while the outer crescent essentially disappears. For the ngVLA wavelength bands, these features become increasingly diffuse and faint at longer wavelengths ($\geq 3.0$ mm) and are nearly undetectable at 10 mm, consistent with reduced dust emission and beam dilution of small-scale structures. Compared to the original unconvolved images, this highlights how observational resolution and sensitivity significantly affect the ability to detect and characterize such fine-scale substructures in vortices. However, our synthetic images do not include noise which is another important component to consider in order to accurately determine if the outer vortex, for example, is resolved or not. We also understand that while the absolute contrast in dust emission may vary with the seeding time, the qualitative morphology, particularly the asymmetric structures associated with the vortices, will remain robust.

\subsection{Brightness contrast and aspect ratio}

There are two main physical properties we can analyze from our disk imaging that can be directly compared with those estimated by ALMA observations from disks with dust crescents. The dust crescents in disks observed in sub-mm by ALMA are identified by their large brightness contrast. We define the brightness contrast ratio as $\frac{I_\mathrm{\nu,max}}{I_\mathrm{\nu,min}}$ of the dust crescent as a function of radial location. Using the same time snapshot as in previous analyses, we calculated the brightness contrast of the brightest inner dust crescent, adopting the $85\%$ peak intensity level for all ALMA bands shown in Figure \ref{fig:radtrans_0incl}, except for Band 9. Because Band 9 exhibits a much lower brightness contrast and appears as a nearly smooth, ring-like structure with little evidence of substructure, we instead used a $95\%$ threshold; using $85\%$ would no longer capture a crescent-shaped feature but rather a continuous ring. We find contrast ratios of 4.23 (Band 6), 3.93 (Band 7), 3.57 (Band 8), and 2.49 (Band 9), indicating that the contrast decreases toward shorter wavelengths. From Band 9 to Band 6, $I_\mathrm{\nu,max}$ drops by $\sim$21$\%$. As shown in Figure \ref{fig:radtrans_0incl}, the contrast between the crescent-shaped features and the surrounding disk increases with observing wavelength, while the overall intensity declines slightly at longer wavelengths, reflecting both the preferential tracing of larger, more concentrated dust grains and the reduced thermal emission.

\begin{figure*}[htp!]
\centering
\includegraphics[width=17.5cm]{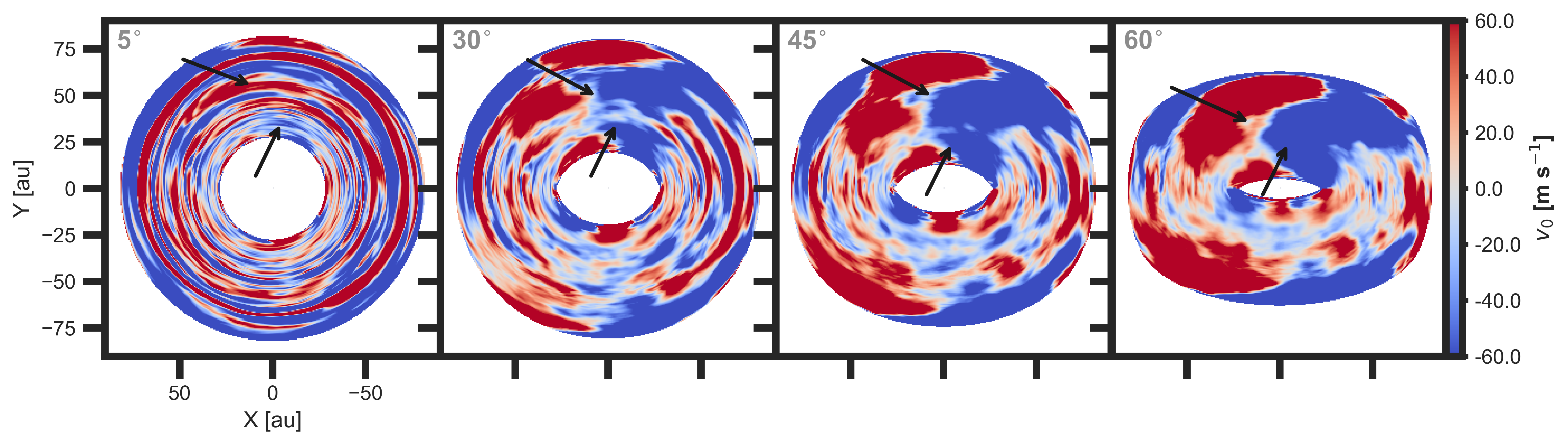} 
\caption{Velocity residuals of the simulated disks at different disk inclinations.}
\label{fig:moment_maps}
\end{figure*}

The second physical property is the aspect ratio, $\chi$ defined earlier as the ratio between the azimuthal extent to the radial extent of the dust crescent, as a function of its radial location. For this, we fit \textsc{EllipseModel()} from the \textsc{skimage.measure} module in Python which uses the least-square geometric fitting method to estimate the parameters of an ellipse from the X and Y pixels points in the image data. The ellipse model uses the spatial distribution of the dust crescent pixels corresponding to 85$\%$ of $I_\mathrm{\nu,max}$ for all bands (and 95$\%$ for Band 9), and computes the best-fit ellipse by minimizing the algebraic distance between the image points and the conic representation of an ellipse. We use the lengths of the semi-major and semi-minor axes as the azimuthal and radial extent for $\chi$. We estimate the aspect ratios for the four ALMA wavelength bands to be: $5.1$ (Band 9), $4.0$ (Band 8), $3.8$ (Band 7), and $3.6$ (Band 6). By dividing the estimated semi-major axis by $\sigma_\mathrm{beam}$, we find that the inner vortex is well resolved in Band 6 (ratio $\sim$3.6) and in Band 9 (ratio $\sim$2.6). Our estimates of the brightness contrast and aspect ratio are further compared and discussed in the context of the observed estimations in Section \ref{sec:dust_cresc_properties}.

\subsection{Effects due to disk inclination}
\label{sec:effect_incl}

So far, we have analyzed the observational features of our model assuming a disk inclination of $0^{\circ}$. In Figure \ref{fig:effect_incl}, we explore how the appearance of the dust crescents changes with increasing inclination. We find that the crescent becomes progressively more elongated and can resemble a ring-like structure at higher inclinations. This projection effect has been reported in previous studies \citep{Doi_2021, Villenave_2025}, where optically thin, vertically extended, and radially narrow dust rings exhibit azimuthal over brightness asymmetries w. In particular, emission is enhanced along the projected semi-major axis because the line-of-sight intersects a longer path through the vertically extended dust layer, while the semi-minor axis samples a shorter vertical column and therefore appears fainter. Similar inclination-dependent azimuthal smearing of dust substructures has also been discussed in recent observational analyses \citep[e.g.][]{Ribas_2024}. As seen in the highest-inclination case in Figure \ref{fig:effect_incl}, this geometric effect can partially wash out crescent contrasts. This result is encouraging, as many disks observed in ALMA Bands 6–7 have moderate to high inclinations, implying that vortex-induced crescents may remain undetected in some systems due purely to projection effects.

We further analyze the radial intensity profiles of the images for all disk inclination cases to identify any potential intensity bumps at the location of the dust crescent. For the $0^{\circ}$ inclination case, which shows the clearest intensity features, the maximum peak corresponds to the location of the pressure bump as in Figure \ref{fig:pressure} (bottom). While the intensity generally decreases with radius, a second peak at the location of the outer vortex is not evident, despite the presence of minor fluctuations. Similarly, the radial intensity profiles for disks at higher inclinations exhibit a smoother distribution.

\subsection{Velocity residuals around the dust asymmetries}

In order to post process the simulation data in the gas and explore observable features to compare with ALMA observations, we use \textsc{RADMC-3D} and perform a line transfer of a line velocity of interest assuming local thermodynamic equilibrium (LTE) conditions, following the methodology presented in \citet{Barraza-Alfaro_2021}. We use the $^{12}$CO $(J=2-1)$ line transition centered at 230.538 GHz. For the molecular number density, we assume a constant fraction of $^{12}$CO relative to H$_{2}$ of $1 \times 10^{-4}$. The radiative properties of the CO molecular data is done with the Leiden LAMDA database\footnote{\url{https://home.strw.leidenuniv.nl/~moldata/}} \citep{Schoier_2005}. We also consider the 3D structure of all velocity components, and assume that the dust temperature is equal to the gas temperature. To generate the synthetic data cubes, we set the velocity range to go from -6 km~s$^{-1}$ to 6 km~s$^{-1}$ within a total of 40 channels, each with a resolution of 0.15 km~s$^{-1}$ in \textsc{RADMC-3D}. 

Figure \ref{fig:moment_maps} displays the moment 1 maps of the velocity residuals, $v_0$, obtained by subtracting the background Keplerian rotation profile from the simulated gas velocity field for the same four disk inclinations shown in Figure \ref{fig:effect_incl}. These maps were computed using the \textsc{BETTERMOMENTS}\footnote{\url{https://github.com/richteague/bettermoments}} Python package \citep{Teague_2018}, which applies a quadratic fitting method centered on the brightest pixel of the point-spread function to more precisely estimate the line centroids of spectral profiles. Once the pure Keplerian velocity is subtracted from the line-of-sight velocity maps, significant deviations from Keplerian rotation emerge across all inclinations. For the lowest inclination (first panel), concentric ring-like features of sub-Keplerian residuals dominate the disk. At the location of the outer vortex (Y = 50 au), a characteristic quasi-dipolar red-blue pattern is evident, indicating faster-moving gas (blue) following the disk’s rotation, and slower gas (red) on the opposite side moving against it. In contrast, Keplerian deviations near the inner vortex are more difficult to discern; however, as the inclination increases, this red-blue pattern becomes increasingly apparent in both vortex regions. Outside the vortices, the concentric ring-like Keplerian deviations remain somewhat symmetric, whereas in the vortex-hosting regions, this symmetry is clearly disrupted. We do not observe clear spiral features in the velocity residuals near the vortex regions. Future studies should analyze the velocity residuals of CO isotopologues to more thoroughly investigate the gas dynamics surrounding vortices.

\section{Discussion}
\label{sec:discussions}

\subsection{Model limitations}

In terms of thermodynamics, the simulation employs a simplified cooling prescription that is short enough to allow the growth of the VSI. We also conducted additional simulation runs using different cooling times to investigate the behavior of vortices in dust trapping under both optimal and marginal VSI conditions; these results will be presented in a forthcoming paper. The most recent state-of-the-art studies addressing the vertical thermal structure in VSI-unstable disks incorporate radiative transfer into radiation hydrodynamics (RHD) models. For instance, \citet{Melon-Fuksman_2024} evolve the frequency-integrated radiation energy flux, while \citet{Zhang_2024} implement angle-dependent (anisotropic) radiative transfer to more accurately model radiation transport in both optically thin and thick regimes. Both studies find that the VSI results in a quiescent midplane, with turbulence primarily confined to the disk’s upper layers. These approaches have been consistently applied in VSI-unstable disk models to explore gas–dust dynamics under varying gas cooling prescriptions, opacities, and molecular emissivities, concluding that VSI growth and stability are most favorable in regions near the disk surface $(Z/R \approx 0.1-0.2)$, where stratification is strongest \citep[i.e.,][]{flock_2021, Melon-Fuksman_2024}. More advanced radiative transfer schemes, such as those used by \citet{Muley_2023}, further resolve the energy exchange between gas, dust, and radiation, underscoring their potential for future applications in modeling hydrodynamic instabilities.

We limit our analysis to processes occurring prior to dust aggregation and the initial formation of planetary embryos. While this study demonstrates that vortices can generate dust asymmetries in sub-mm continuum emission within turbulent protoplanetary disks, further VSI simulations are necessary to investigate the longevity of these vortices under the influence of dust backreaction. The formation of planetesimals is favored in scenarios where the SI develops following the onset of the VSI in an isothermal configuration, with strong clumping occurring for dust concentrations as low as 0.5$\%$ (St = 0.1) or 1$\%$ (a = 3 mm) \citep{Schafer_2020, Schafer_2022}. However, by changing the equation of state, gas buoyancy can affect the VSI in the disk midplane \citep{Lin_2017, Lin_2019}. 

\subsection{Resolution effect}
\label{sec:resolution_effect}

It is worth noting that a recent high-resolution study by \citet{Shariff_2024} and \citet{Lesur_2025}, reaching up to 70 and 250 cells per scale height, respectively, reported no vortices in the midplane of their VSI simulations using an effectively isothermal setup. Instead, these simulations only show small-scale, isotropic fluctuations below the disk scale height. They suggest that this absence of vortices could be linked to the elliptical instability \citep[E.I.;][]{Lesur_2009}, not present in lower resolution runs of previous works with 18-50 cells per H; e.g., \citealt{Manger_2018, Richard_2016, Huang_2025b}), including the present work (37 cells per H). We note that the simulations in \citet{Lesur_2025} report evolution in inner orbits measured at 
$R=0.25 R_{0}$, which corresponds to $<$187 local orbits for the longest resolution which is significantly shorter than the timescale over which vortices emerge in our models. Regarding the resolution used in \citet{Shariff_2024}, it is difficult to directly compare their resolution with ours because they employ a Pad\'e filter, a high-order numerical smoothing operator that selectively damps unresolved small-scale fluctuations near the grid scale while preserving the resolved large-scale flow; its operation also requires an accompanying shock-viscosity treatment. We encourage future studies to investigate the occurrence of long-lived vortices in high-resolution models across a wide range of cooling timescales, since vortices may persist at longer cooling timescales (\citealt{Richard_2016}, Sengupta et al. in prep.)

We investigate the kinetic energy spectrum of our simulations by extracting a smaller box region of the domain centered at 1.25 in code units, taking $r \in[1.1,1.4]$ in code units, and $\theta \in [1.57,1.81]$, and $\phi \in [3.14,3.68]$ in radians. To perform a Fourier transform (FFT) on the simulation data, the grid points must be equally spaced so that the transformed fields satisfy the periodicity assumptions of the FFT. Because the radial grid in our case is non‑uniform, we first resample it onto a uniform grid using linear interpolation. We then apply a Hanning window to the selected box to reduce edge effects in case the boundaries are not periodic. After windowing, we multiply the simulation variables by the window and compute the Fourier transform of the momentum components in each direction, followed by the Fourier transform of the velocities. The resulting energy spectrum is defined as the sum of each velocity Fourier component multiplied by the complex conjugate of its corresponding mass‑flux Fourier component. The 3D spectrum is obtained by binning the Fourier‑space energy into spherical shells using evenly spaced wavenumber bins, summing the total energy, $E(k)$, in each shell, dividing by the shell thickness to obtain an energy density, and finally normalizing the spectrum so that its integral over all wavenumbers equals unity. This methodological procedure follows the approach described in Appendix B of \citet{Melon-Fuksman_2024}.

From Fig. \ref{fig:energy_spectrum} we observe the energy spectrum together with two reference slopes: a Kolmogorov-like power-law slope (green) and a much steeper power-law slope (red). The Kolmogorov-like regime is indicative of a turbulent cascade; however, in anisotropic, rotating, and vertically stratified flows such as VSI turbulence, this slope does not uniquely imply a forward cascade from large scales. Energy may instead be injected near the fastest-growing or marginal VSI wavelengths and redistributed both upscale and downscale. By estimating the scale-dependent Rossby number, we find that $\mathrm{Ro}<0.1$ near the spectral kink and decreases toward both larger and smaller scales, indicating that rotation increasingly constrains the flow away from the injection scale. This behavior is consistent with quasi-two-dimensional rotating turbulence, in which energy is preferentially transferred upscale while enstrophy cascades downscale, naturally explaining the coexistence of a Kolmogorov-like slope and a steeper high-$k$ regime \citep[see also][]{Manger_2018}. Here $k=2\pi/\lambda$ denotes the Fourier wavenumber associated with spatial scale $\lambda$. The two vertical gray lines mark the characteristic VSI wavelengths in the Fourier space, expressed in units of $kH$. Structures with wavelength equal to the disk scale height ($\lambda = H$) appear at $kH = 2\pi$, while the fastest‑growing and marginal modes located at $kH = 40 $ (dotted-line) and $kH = 20$ (dashed-line), respectively. We theoretically calculate the fastest‑growing VSI wavelength using the prescription from \citet{Umurhan_2016}, further supported by the numerical analysis of \citet{Shariff_2024}, where $\frac{\lambda}{H} = \pi |q|(\frac{H}{R})^{2}$ = 0.16. Using the local scale height at the spectrum center, we obtain $\lambda=0.020$ with the marginal mode at $2\lambda$. Converting to Fourier space, this yields the characteristic wavenumbers marked in vertical gray lines in Fig. \ref{fig:energy_spectrum}. These characteristic scales lie close to the transition toward the dissipation range, where the spectrum steepens sharply. This proximity suggests that part of the VSI-injected energy may be removed numerically rather than fully cascading through a resolved inertial range, although the presence of both Kolmogorov-like and steeper slopes indicates a turbulence regime strongly influenced by shear, rotation, and vertical stratification.

By multiplying $\frac{\lambda}{H}$ by the number of cells per scale height shows that the fastest‑growing modes are resolved within roughly 6 grid points, meaning that the power injected at that spatial scale can accumulate and contribute to the generation of the vortices. The fact that the energy injection occurs close to the dissipation range implies that VSI energy is dissipated locally and turbulence remains underdeveloped at small scales (large $k$). However, in our case the vortices remain self‑sustaining due to continuous VSI driving throughout the simulation time. Thus, early damping of VSI motions effectively prolongs vortex survival relative to what would occur in a fully resolved turbulent cascade.

\begin{figure}[htp!]
\centering
\includegraphics[width=7.5cm]{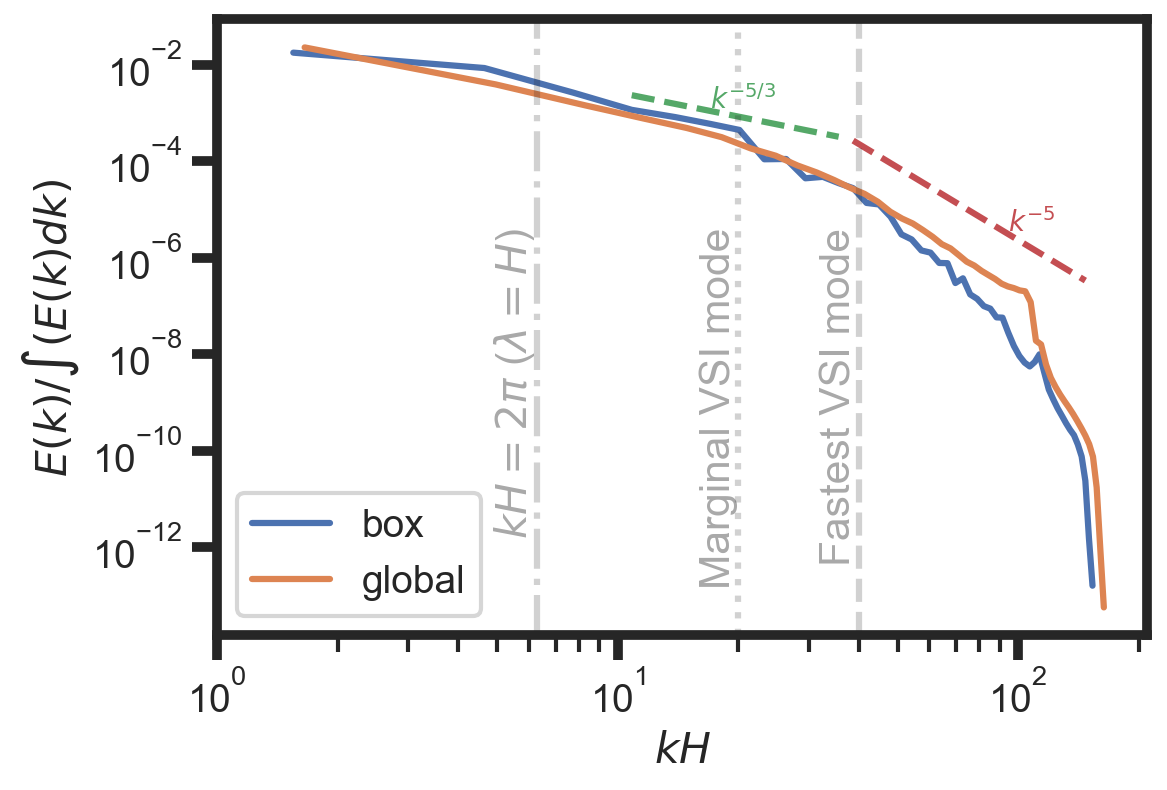} 
\caption{Energy spectrum for a small box extracted from the simulation domain (blue curve) and for the full domain (orange curve) at 400 orbits. Overplotted are two power laws, $k^{-5/3}$ and $k^{-5}$ illustrating the Kolmogorov regime and a steeper dissipation regime, respectively. The dotted and dashed vertical gray lines mark the wavenumbers corresponding to the marginal and fastest‑growing VSI modes, based on the prescription from \citet{Umurhan_2016}. The dash-dotted vertical line denotes to the mode whose wavelength equals the local disk scale height $\lambda=H$, corresponding to $kH=2\pi$.}
\label{fig:energy_spectrum}
\end{figure}

\subsection{Dust vertical diffusion}
\label{sec:dust_vert_diffusion}

\noindent The VSI drives turbulent vertical gas motions that play a key role in determining the extent of dust vertical diffusion in the disk. We estimate the degree of vertical stirring, $\alpha_{z}$, from the directly fitted value of $H_{d}^\mathrm{}/R_\mathrm{50au}$ obtained from the dust grain distribution (Figure \ref{fig:dust_gauss}) and the Stokes number for all grain sizes at the midplane (see Appendix \ref{sec:stokes}). The vertical stirring is defined as

\begin{equation} 
\label{eq:alphaz}
\alpha_{z} = \mathrm{St}  \left [ \left( \frac{H}{H_{d}} \right )^{2} -1 \right]^{-1},
\end{equation}

\noindent \citep{Dubrulle_1995} yielding $\alpha_{z} = 1.7\times 10^{-3}$ with St$=1.4\times10^{-2}$ for the 1 mm grains, $\alpha_{z} = 1.3\times 10^{-3}$ with St$=7.1\times10^{-3}$ for the 500 $\mu$m grains, and $\alpha_{z} = 6.4\times 10^{-4}$ with St$=1.8\times10^{-3}$ for the 100 $\mu$m grains. Because dust backreaction is absent, the gas turbulence is identical for all grain sizes; however, the inferred $\alpha_{z}$ values reflect how each Stokes number responds to that same flow. In this sense, $\alpha_{z}$ should be interpreted as a grain‑dependent measure of vertical stirring efficiency rather than a unique turbulent viscosity of the gas. We note, however, that the Stokes number entering Eq. (\ref{eq:alphaz}) implicitly depends on the characteristic turbulence correlation or eddy turnover timescale, $\tau_{\rm eddy}$. In VSI turbulence this timescale is not well constrained. The commonly adopted definition $\mathrm{St}=t_{\rm stop}\Omega_k$ assumes that the eddy turnover time is of order $\Omega_k^{-1}$. More generally, $\mathrm{St}=t_{\rm stop}/\tau_{\rm eddy}$, such that variations in $\tau_{\rm eddy}$ would modify the inferred stirring efficiency. Therefore, the $\alpha_z$ values derived here should be regarded as effective diagnostics under this standard assumption rather than absolute measures of the intrinsic gas diffusivity.

The dust scale heights we measure (see Section \ref{sect:dust_dist} and Figure \ref{fig:dust_scale_height_over_time}) are comparable to the vertical wavelengths of the VSI modes. This is consistent with the nonlinear VSI dynamics described by \citet{Latter_2018}, who showed that vertical roll‑up of VSI modes produces coherent overturning motions capable of vertically stirring dust on these same scales.

Our estimate of the dust scale height, $H_{d}^\mathrm{1mm}/R_\mathrm{50au}=0.032$, is very similar to the value reported in RHD VSI simulations by \citet{Flock_2020}, $H_{d}=0.034$, for the same grain size. Consistently, we obtain $\alpha_{z}$/St = 0.12 for the 1 mm particles, close to the value of 0.10 reported by \citet{Flock_2020}, indicating that turbulent stirring dominates over vertical settling in both cases. Since the VSI strength is highly sensitive to initial disk conditions and the adopted thermodynamic treatment—particularly the cooling times with radius \citep[e.g.,][]{Richard_2016, Manger_2018, Pfeil_2021}—further exploration of dust distributions under different initial setups is needed.

\citet{Villenave_2025} proposed that the VSI is not active in disks with well-settled dust scale heights, corresponding to $\alpha_{z}$/St $\leq 0.1$, a threshold motivated by the finding of \citet{Stoll_2017} that $\alpha_{z}=650\alpha_{r}$ under isothermal conditions. In 2D isothermal VSI simulations, \citet{Flores-Rivera_2025}—using the same disk profile $(p=-1.5, q=-1)$ as \citet{Stoll_2017}—found a dust scale height of $H_{d}=0.033$ for 1 mm particles, with $\alpha_{z}$/St $\approx 0.12$, consistent with a regime of efficient VSI-driven mixing. This agrees with the trend shown in Figure 5 of \citet{Villenave_2025}, where $\alpha_{z}$/St values below 0.1 correspond to weak vertical stirring and diminished VSI activity. Notably, their fits were obtained for 88 $\mu$m grains, comparable to the 100 $\mu$m case in our study, for which we find $\alpha_{z}$/St $\approx 0.36$, again in line with their results. By contrast, \citet{Flock_2020}, who included radiative transfer in their VSI simulations, reported $\alpha_{z}$/St ratios closer to the marginal threshold proposed by \citet{Villenave_2025}. These comparisons highlight that VSI strength is highly sensitive to both disk structure and thermodynamic treatment, underscoring the need for further simulations with dust that systematically explore a wider range of initial conditions and realistic physics to refine this threshold.

We further investigate the eddy timescale governing particle vertical diffusion. This analysis also provides an empirical estimate of the turbulence correlation time, which is otherwise uncertain in VSI-driven flows and implicitly assumed in the Stokes number definition used above. From the vertical settling-diffusion equilibrium definition \citep{Dubrulle_1995, Youdin_2007Icar} 

\begin{equation} 
\label{eq:hd_h}
\left( \frac{H_{d}}{H} \right )^{2} = \frac{D_z}{\mathrm{St} c_s H}
\end{equation}

\noindent where $D_z$ is the dust vertical diffusion related to the eddy turnover time defined as \citet{Fromang_2006}

\begin{equation} 
\label{eq:tau_eddy}
D_z = \langle v_z^{2} \rangle \tau_\mathrm{eddy} \rightarrow \tau_\mathrm{eddy} \Omega_k = \frac{\mathrm{St} c_s^2}{\langle v_z^{2} \rangle} \left( \frac{H_{d}}{H} \right )^{2}
\end{equation}

Outside the vortices, we evaluate the averaged vertical velocity of particles over the radial range 40–50 au, time-averaged after 300 orbits. For 1 mm grains, we use the averaged Stokes number between $(Z/H_{R})=0$ and $(Z/H_{R})=1$ which is 0.0185 and the vertical velocity of the particle is $\langle v_{z}^{2} \rangle / c_{s}^{2} = 7.2\times 10^{-4}$, the estimated eddy timescale for a particle inside a vortex is $\tau_\mathrm{eddy}\Omega_{k} = 2.63$. For comparison, \citet{Stoll_2016} reported a value of 0.2, while \citet{Flock_2020} obtained 20 based on $\alpha_{z} = 5.4\times10^{-3}$ and $\langle v_{z}^{2} \rangle / c_{s}^{2} = 2.8\times10^{-4}$. Our results are consistent with the theoretical framework of \citet{Sengupta_2024}, who argued that anisotropic turbulence naturally produces Schmidt numbers different from unity. By inferring the eddy turnover time directly from the dust-diffusion properties in our VSI simulations, we obtain $\tau_\mathrm{eddy} \Omega_{k}$$\sim$ 1-3, placing our system the $\Omega_{L}>\Omega_{k}$ regime identified in their work.

Inside the vortex, the vertical diffusion (equation \ref{eq:alphaz}) is estimated to be $\alpha_\mathrm{z,vortex} = 5.8\times10^{-4}$ using a dust scale height of $H_{d}/R=0.02$ or $(H_{d}/H)^{2}=0.04$ (see Figure \ref{fig:dust_scale_height_over_time}). The corresponding eddy timescale is 1.17, calculated from a value of $\langle v_{z}^{2}  \rangle / c_{s}^{2} = 6.3\times 10^{-4}$ (Figure \ref{fig:veloc_profile}). The $\alpha_\mathrm{z,vortex}$ value is a factor of $\sim$3 times lower than previously determined, more global value of $\alpha_{z} = 1.7\times 10^{-3}$ for the 1 mm grain size case. These very low values suggest that vortices preferentially enhance diffusion in the radial and azimuthal directions rather than vertically.

\subsection{Dust crescents and disk properties}
\label{sec:dust_cresc_properties}

Dust crescents are readily identifiable in sub-mm observations due to their characteristic azimuthal brightness variations within the dusty ring. To date, a total of 19 dust crescents have been identified across 13 disks \citep{Bae_2023}, suggesting that they may be relatively rare features in protoplanetary disks. However, as shown in Figure~\ref{fig:effect_incl}, when exploring crescent morphology under varying disk inclinations, we find that the apparent asymmetry of the dust crescent changes significantly with increasing inclination due to geometric projection effects discussed in Section~\ref{sec:effect_incl}. At high inclinations, the crescent-like morphology can become so distorted that it no longer resembles a crescent. Additionally, brightness contrast varies with inclination; higher inclinations reduce the line-of-sight interception of photons through the semi-minor axis, making that region appear fainter. We also predict that dust crescents traced by large particles in the disk midplane are unlikely to be significantly affected by scattering phase functions, as scattering is not expected to play a major role at such depths.

Given the limited number of dust crescents identified in the literature, little is known about how their size and brightness depend on global disk properties. The aspect ratio of dust crescents is thought to be a key parameter for constraining the physical mechanisms responsible for their formation. Although our estimates of the aspect ratio at $0^{\circ}$ disk inclination fall within the range reported for observed dust crescents at similar radial locations \citep[see Figure 5 in][]{Bae_2023}, both the radial widths (ranging from 7 to 80 au) and azimuthal extents (40 to 190 au) exhibit significant variation. Consequently, there appears to be no clear correlation between aspect ratio and radial location, especially when projection effects due to disk inclination introduce further degeneracies. \citet{Bae_2023} also report that the occurrence rate of crescents is approximately 30$\%$ for disks with $M_\mathrm{disk}/M_{\odot} < 0.05$. Our simulated aspect ratios are consistent with this trend, as our model disk has a mass of 0.038 $M_\odot$. Regarding crescent multiplicity, three disks to date—MWC 758 \citep{Dong_2018}, HD 143006 \citep{Benisty_2018}, and HD 139614 \citep{Muro-Arena_2020}—have been found to host more than one crescent, consistent with our predictions that multiple vortices can coexist in a single disk.

Other works by \citet{Flock_2020} and \citet{Blanco_2021} have explored and characterized the observational predictions and substructures in the distribution of mm-sized and small particles in 3D RHD models of the VSI. Both studies demonstrate that a vortex generated by the VSI can trap dust and create a localized azimuthal bump in the dust continuum emission, particularly at the inner edge of a planet-induced gap. The resulting dust concentration is more compact and produces a higher brightness contrast for larger grains, leading to stronger emission features at longer wavelengths—consistent with our findings. Additionally, they found that VSI-induced dust crescents exhibit radial widths of $\sim$2 au, which may serve as observational signatures of the VSI in protoplanetary disks.

Whether arising from the RWI or the KH instability, vortices are expected to be ubiquitous in both VSI-active and non-VSI disk environments. To guide observers, we propose that high-angular-resolution observations ($\geq$ 14 au, as in the exoALMA program) of velocity residuals can help distinguish between vortices triggered by planets and those arising independently. If spiral features are detected in velocity residuals or in NIR scattered-light images, the RWI vortex is likely excited by a giant planet carving a gap. In contrast, in systems where spirals are absent—such as HD 143006, HD 34284, or the outer vortex in MWC 758—the vortices are, perhaps, less likely to be planet-induced.

\subsection{Dust-to-gas feedback effect}
\label{sec:dust_t0_gas_effect}

There have been relatively few studies exploring the numerical effects of dust-induced feedback on gas dynamics within vortices. \citet{Johansen_2004} provided the first three-dimensional demonstration that anticyclonic vortices can efficiently trap dust in protoplanetary disks, highlighting their potential role as long-lived sites of planetesimal formation and paving the way for connecting vortex theory with future ALMA observations. \citet{Meheut_2012a} notably found that the vertical velocity structure within a vortex differs at higher disk altitudes. In their simulations, small particles ($<2$ cm) experienced vertical lifting due to a positive feedback effect on the vertical velocity field. \citet{Fu_2014} investigated dust back-reaction in a planet-induced vortex and found that when the local dust-to-gas mass ratio approached unity, a dynamical instability could be triggered within the vortex. This instability altered the vorticity and led to a reduced vortex lifetime. In contrast, \citet{Surville_2016}, using very high-resolution simulations ($N_{r} \times N_{\phi} = 2048 \times 4096$), concluded that vortices could survive up to 1000 orbits even when accounting for dust back-reaction, eventually transitioning into ring-like structures that may become susceptible to the SI and enable rapid planetesimal formation \citep[see also][]{Raettig_2015}. Similarly, \citet{Miranda_2017} performed HD simulations including dust feedback and produced synthetic observations, finding that while dust feedback may weaken vortices and slow dust accumulation, the dust crescents remain detectable for thousands of orbits, unlike in the planet-induced vortex scenario. They attribute this to the continual mass accumulation at the gap edge, which replenishes the vortex and prevents its dissipation. More recently, \citet{Huang_2025} conducted 3D high-resolution VSI simulations (with ~230 cells per scale height) and found evidence of dust clumping, with approximately 2$\%$ of the dust mass trapped in vortices potentially undergoing gravitational collapse. They also identified small-scale dust clumps forming in azimuthal vortices within shearing zonal flows, which may be associated with Kelvin–Helmholtz (KH) instabilities. These findings emphasize that VSI-driven turbulence is not purely diffusive, but instead enables interactions with multiple hydrodynamic instabilities. In our work, we similarly confirm the interplay of three key instabilities: the VSI, which dominates the global gas dynamics; the RWI, which supports long-lived, large-scale vortices; and the KH instability, which emerges at the shear interfaces of zonal flows.

\subsection{Implications to planet formation}

Vortices are widely considered prime sites for planetesimal formation due to their efficiency in trapping pebbles, thereby enhancing the local dust-to-gas mass ratio and enabling the growth of planetary embryos \citep{Johansen_2004, Lyra_2008a}. \citet{Lyra_2009a} conducted global 2D simulations incorporating self-gravity and found that solid material accumulated within vortices and subsequently grew through collisions and pebble accretion \citep{Lambrechts_2012}, ultimately can form Earth-mass planets. Later, \citet{Lyra_2009b} demonstrated that vortices could be sustained in 3D through the RWI at gap edges where a viscosity transition occurs due to a Jupiter-mass planet opening a gap. When a Jupiter-mass planet grows over a sufficiently long timescale (e.g., >700 orbits), the vortex triggered at the gap edge can become highly elongated (extending over approximately $180^{\circ}$ in azimuth), which appears to be a promising characteristic for distinguishing dust asymmetries produced by a vortex formed due to planet-induced gap opening \citep{Hammer_2017, Hammer_2019}. The persistence of RWI-induced vortices in three dimensions was subsequently confirmed by multiple studies \citep{Meheut_2010, Meheut_2012a, Meheut_2012b, Meheut_2012c, Lin_2012a, Lin_2012b, Lin_2013, Lyra_2015}. In our case with no self-gravity, the vortices are able to trap a fraction of the Moon mass around their center.

Vortices offer a compelling solution to the drift barrier in planet formation \citep{Birnstiel_2010}. While micrometer-sized dust particles can coagulate into millimeter-sized pebbles, further growth is inhibited by rapid radial drift toward the central star. This drift, dependent on particle size \citep{Weidenschilling_1977}, arises because gas in the disk, supported by pressure, rotates at sub-Keplerian speeds, causing dust particles to experience drag and spiral inward. However, since vortices modify the local pressure distribution in the disk by creating pressure maxima, they act as dust traps that draw grains toward their centers of rotation, preventing them from drifting inward and being lost. The dust density enhancement at the vortex \citep{Barge_1995, Klahr_2006, Meheut_2012a} provides favorable conditions for gravitational collapse or the streaming instability \citep{Raettig_2015, Raetting_2021}, to ultimately promote planetesimal formation. 

\subsection{Vortices as a storage reservoir for CV chondrites}

In cosmochemistry, the initial solid inventory of the protoplanetary disk is represented by chondritic meteorites—fragments of primitive, undifferentiated asteroids primarily composed of chondrules and fine-grained dust that formed in other regions of the disk before being accreted onto the parent body \citep{Davis_2014}. The accretion ages of these chondrites are determined by modeling the thermal histories of their parent bodies, driven by heating from the decay of $^{26}$Al \citep{Grimm_1993}. These thermal profiles are compared with the formation ages and peak temperatures of secondary minerals such as fayalite (Fe$_{2}$SiO$_{4}$) and complex carbonates ([Ca,Mg,Mn]CO$_{3}$).
These minerals serve as valuable indicators of aqueous alteration because they form through fluid-assisted reactions on the parent bodies, thereby providing a lower limit for the accretion age of the chondrites. Using the Mn-Cr dating method \citep{Doyle_2015}, the formation age of fayalite in CV chondrites has been determined to be 4.2 $\pm$ 0.8 million years after the formation of CV Calcium-Aluminium-Inclusions (CAIs). These CAIs mark the reference time ($t_{0}$) for the first solids in the solar system, as established by the U-corrected Pb-Pb dating method \citep{Connelly_2012}.
Based on this lower limit, the modeled accretion age of the CV chondrite parent body is approximately 2.5 million years after CAI formation. This time gap between solar system formation and the accretion of CV parent bodies aligns with the timescales over which particles can be retained in vortices before dissipation. We therefore propose that vortices acted as storage reservoirs, accounting for both the preservation of early solar system materials (CV CAIs) and the delayed accretion of CV chondrules into their parent bodies.

\section{Conclusions}
\label{sec:conclusions}

We studied the 3D dynamics of dust in turbulent protoplanetary disks, examining how particles are trapped and retained within vortices self-sustained by the Vertical Shear Instability (VSI). We performed three-dimensional simulations using Lagrangian particles of three different sizes: 1 mm, 500~$\mu$m, and 100~$\mu$m, to assess how vortices capture particles of varying sizes. Our main findings are summarized below:

\begin{enumerate}
      \item We show that vortices can trap dust particles in disks dominated by VSI-driven turbulence. Our simulations show the formation of multiple vortices, including small-scale structures likely induced by shearing azimuthal flows from the Kelvin-Helmholtz instability. Two large-scale vortices become fully developed and stable after approximately 250 orbits. The inner vortex efficiently captures most of the 1 mm particles, that underwent rapid radial drift and accumulate at the inner pressure maxima. Similarly, the 500~$\mu$m particles are trapped by both large-scale vortices. In contrast, the 100 $\mu$m particles are not captured by the inner vortex, as its efficient trapping of mm-sized particles occurs more rapidly, whereas the smaller grains are more dispersed in the surrounding area. This may explain their absence in near-infrared (NIR) scattered light observations. However, these smaller particles are more effectively trapped by the outer vortex, which initially formed at the outer edge of the disk and slowly migrated inward; as a result, it lacks an associated gap in its vicinity.
      \item We find that the VSI drives significant vertical diffusion, with our estimated $\alpha_{z}$ values for 1 mm, 500 $\mu$m, and 100 $\mu$m dust grains consistent with previous work. The ratio $\alpha_{z}$/St$\approx$0.12 for 1 mm grains suggests a regime of efficient turbulent mixing. Inside the vortex, we found the $\alpha_{z}$ to be about 3 times lower than globally determined $\alpha_{z}$ consistent with a low eddy timescale suggesting that vortices preferentially enhance diffusion in the radial and azimuthal directions rather than vertically.
      \item We do not observe spiral structures associated with the two large-scale vortices in the gas velocity residuals or in the distribution of any particle species. Such features are more commonly linked to planet formation in a gap, where spiral wakes are generated in the disk. However, when analyzing the perturbed gas pressure, a trailing spiral-like feature becomes apparent at both the disk midplane and the disk surface.
      \item By analyzing particle trajectories within vortices, we find that turbulent vertical motions driven by the VSI are reduced by up to a factor of two compared to regions where VSI turbulence is fully developed. This reduction implies that particle collision velocities remain low enough to prevent fragmentation, enabling growth through sticking and coagulation. Consequently, particles may reach sizes conducive to gravitational collapse and the formation of planetary embryos, consistent with \citet{Carrera_2025A}, who demonstrated that pebbles drifting into vortices experience reduced turbulence via their drag force on the gas, thereby facilitating further growth. Future simulations should incorporate dust growth processes and dust backreaction to further assess the longevity of vortices and their role in planetesimal formation.
      \item At high disk inclinations, dust crescents are affected by projection effects, which distort their asymmetry and reduce the brightness contrast along the dusty ring. This is important to consider, as many disks observed with ALMA have inclinations greater than $30^{\circ}$, making it challenging to identify dust crescents formed by vortices. Therefore, more than the 13 disks with observed dust crescents \citep{Bae_2023} may in fact host dust vortices. Whether the associated RWI vortices are formed by the VSI or by planetary companions, the presence of spirals observed in NIR scattered light in some systems (e.g., MWC 758 and HD 135344B) suggests that planetary perturbations may also play a role.
\end{enumerate}

In this work, we demonstrate that vortices may serve as initial sites of dust trapping. These could eventually lead to the formation of the first generation of planetesimals and planetary embryos in protoplanetary disks. Recent high-resolution studies \citep{Lesur_2025, Shariff_2024} have further highlighted the sensitivity of VSI-driven structures to numerical resolution, suggesting that small-scale turbulence, rather than coherent vortices, may dominate under effectively isothermal conditions. This emphasizes the need to explore the non-linear regime of the VSI across a broader range of resolutions and cooling timescales. Future work should therefore include significantly higher-resolution simulations to determine whether the apparent absence of vortices at high resolution reflects a genuine physical effect or a limitation of spatial resolution.

\begin{acknowledgements}
We thanks the referee for the helpful discussions that improved the quality of the manuscript. L. Flores-Rivera and M. Lambrechts acknowledge this work is funded by the European Research Council (ERC Starting Grant 101041466-EXODOSS). A.J. is funded by the Carlsberg Foundation (Semper Ardens: Advance grant FIRSTATMO). We also thank Marcelo Barraza, for the helpful discussion about the velocity residuals. To Elishevah van Kooten and Jean Bollard for their input regarding to the application to cosmochemistry.
\end{acknowledgements}

\bibliographystyle{aa}
\bibliography{VSI_II}

\begin{appendix}

\section{Vortex criterion}
\label{sec:criterion}

\begin{figure}[htp!]
\centering
\includegraphics[width=7.5cm]{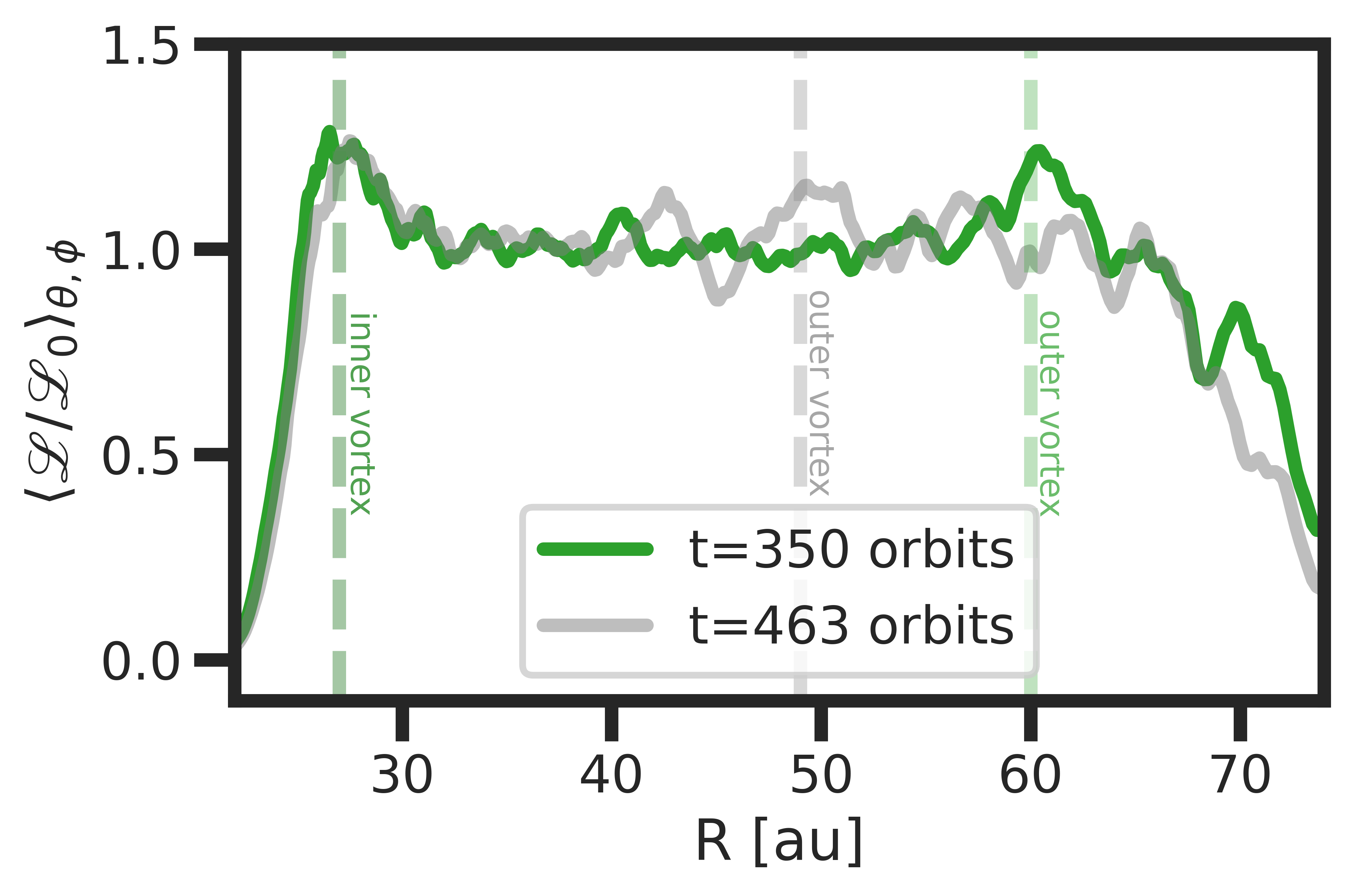}
\includegraphics[width=7.85cm]{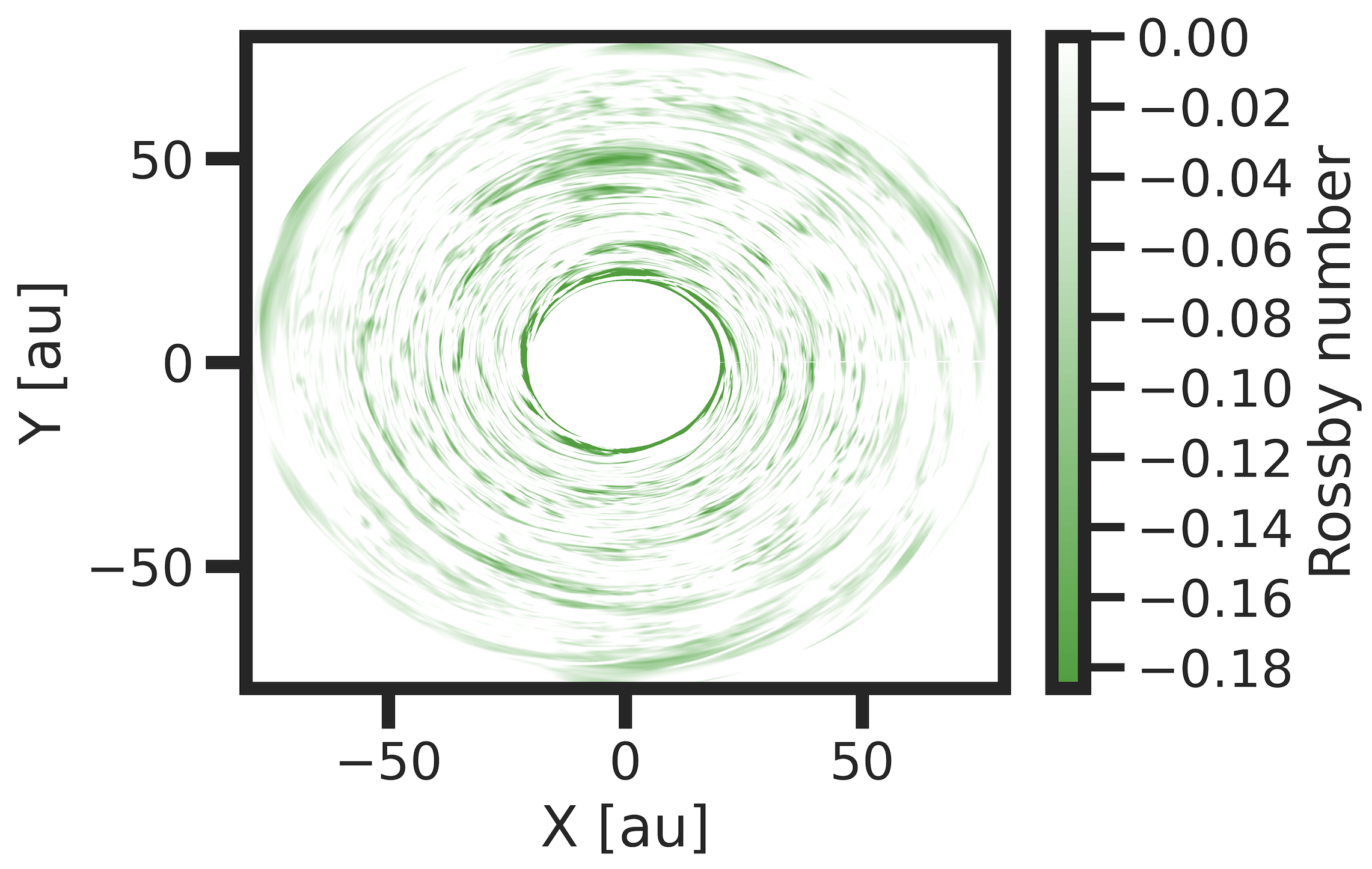}
\caption{Rossby instability criteria at the extrema in the disk.}
\label{fig:rossby}
\end{figure}

The Rossby Wave Instability (RWI) was first derived by \citet{Lovelace_1999} and later expanded upon by \citet{Li_2000, Li_2001} as a global instability that arises at local extrema in pressure or vortensity. It operates in regions where the function $\mathscr{L}$ exhibits an extremum. This critical function, which serves as the criterion for the onset of the RWI, is defined as

\begin{equation}
\label{eq:rossby}
\mathscr{L} = \frac{\Sigma_\mathrm{gas}}{\omega_{z}} \left(
\frac{P}{\Sigma_\mathrm{gas}^{\gamma}} \right)^{2/\gamma}
\end{equation}

where $\Sigma_\mathrm{gas}$ is the surface gas density, $\omega_{z}$ is the vertical component of the vorticity, $P$ is the vertically integrated gas pressure, and $\gamma = 1.4$ is the adiabatic index. We compute $\mathscr{L}$ using azimuthally and vertically averaged profiles of density, pressure, and velocity and normalized it with their initial profiles. Figure \ref{fig:rossby} (left) shows the RWI criterion at at 350 and 463 orbits. The $\mathscr{L}/\mathscr{L}_0$ curve displays multiple global extrema, indicating that large regions of the disk are susceptible to the RWI. Two prominent extrema are identified, and they align radially with the locations of the two large-scale vortices (indicated by vertical dashed gray lines).

The Rossby number, $\mathrm{Ro}$, can also be used to identify regions exhibiting anticyclonic rotation. It quantifies the ratio of Coriolis to inertial forces in the Navier–Stokes equations and is defined as $\mathrm{Ro} = (\nabla \times \vec{v} - \vec{v}_k)_z / \Omega_k$, where the velocity field is measured relative to Keplerian motion and averaged vertically. Figure \ref{fig:rossby} (right) presents the spatial distribution of the Rossby number, highlighting azimuthally elongated vortical structures with negative Rossby numbers. These localized minima signal the presence of anticyclonic vortices, where the dynamics are dominated by a balance between pressure and Coriolis forces.

\section{Vorticity residual for a different model}
\label{sec:vort_residual_model3}

\begin{figure*}[htp!]
\centering
\includegraphics[width=15.2cm]{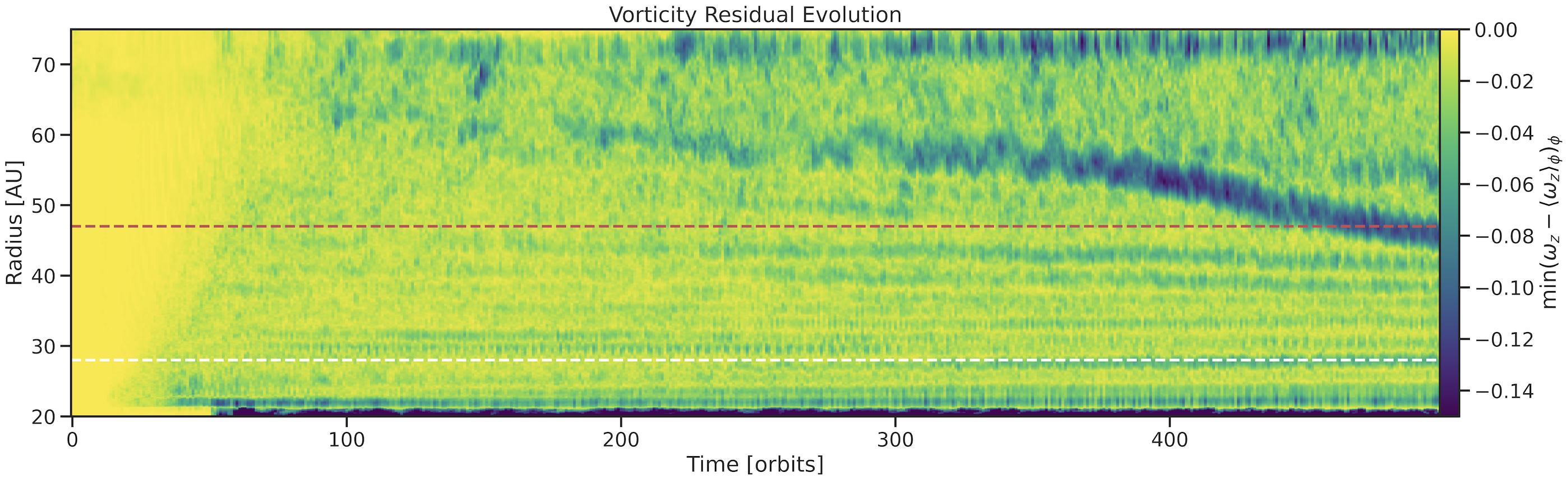}
\caption{Time evolution of the minimum vorticity for a model with identical physical disk parameters but a modified cooling time, $t_\mathrm{cool} = 5\times10^{-3}\Omega_{k}^{-1}$.}
\label{fig:vort_resi_model3}
\end{figure*}

We also present results from a different simulation, with the same physical disk parameters as presented in the main text, but a different cooling time, $t_\mathrm{cool} = 5\times10^{-3} \Omega_{k}^{-1}$ (Flores-Rivera et al., in prep). Applying the same methodology used to construct the vorticity residual over time as in Figure \ref{fig:vort_residual}, Figure \ref{fig:vort_resi_model3} reveals multiple vortex trails at this higher cooling rate. We observe multiple vortices originating between 40 au and 60 au, far from the boundary conditions; this suggests that such features form spontaneously in VSI-dominated disks simulated at the same resolution as in the present case. A caveat remains that at higher resolutions, with less numerical diffusion, the elliptical instability may inhibit vortex formation \citep{Lesur_2009, Lesur_2025}.

\section{Stokes number in the disk}
\label{sec:stokes}

The Stokes number at the midplane ($Z/H_{R} = 0$) is St = 1.4$\times10^{-2}$ for a = 0.1 cm at 50 au. At one scale height ($Z/H_{R} = 1$), the Stokes number increases to St = 2.3$\times10^{-2}$ for the same particle size. This implies that a one-scale-height variation leads to a change in Stokes number by less than a factor of two. Similarly, the Stokes number for particles of size 0.05 cm and 0.01 cm at 50 au is 7.1$\times10^{-3}$ and 1.4$\times10^{-3}$ at the midplane, respectively. This is further supported when considering a turbulent fragmentation velocity of 1 m~s$^{-1}$ and $\alpha = 3 \times 10^{-4}$ (adopted from our VSI turbulence convergence value, see Fig. \ref{fig:alpha_all}  top), which yields a fragmentation-limited Stokes number of 1.8$\times10^{-3}$, corresponding to a particle size of $\simeq$0.02 cm at the midplane at 50 au \citep[see Eq. 33 in][]{Birnstiel_2024}.

These fragmentation-limited Stokes number values apply to the outer vortex region that is centered at $\sim$48.5 au, where the estimated $\alpha$-viscosity ranges from $\sim7 \times 10^{-5}$ from the midplane up to $2~H$. For the inner vortex, the $\alpha$-viscosity is similar ($3$–$7 \times 10^{-4}$ over the same height range), but differences in scale height and sound speed affect the Stokes number. We find that the fragmentation-limited Stokes number in the vortices is $8\times10^{-3}$ for $\alpha_\mathrm{vortex,outer} = 7 \times 10^{-5}$ for the outer vortex and, for the inner vortex it is one order of magnitude lower using $\alpha_\mathrm{vortex,inner} = 7 \times 10^{-4}$.

\begin{figure*}[htp!]
\centering
\includegraphics[width=17.5cm]{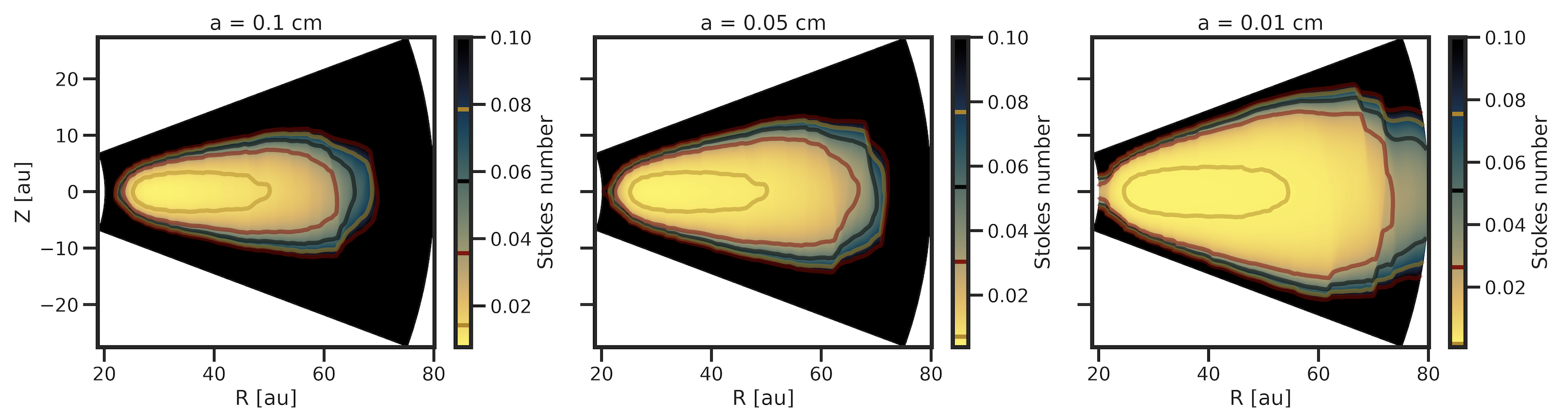}
\caption{Stokes number in the disk. The different colored contour lines in each panel represents different Stokes number at different heights in the disk.}
\label{fig:stokes}
\end{figure*}

\section{Pressure fluctuations around vortices}
\label{sec:more_properties}

\begin{figure*}[htp!]
\centering
\includegraphics[width=7.2cm]{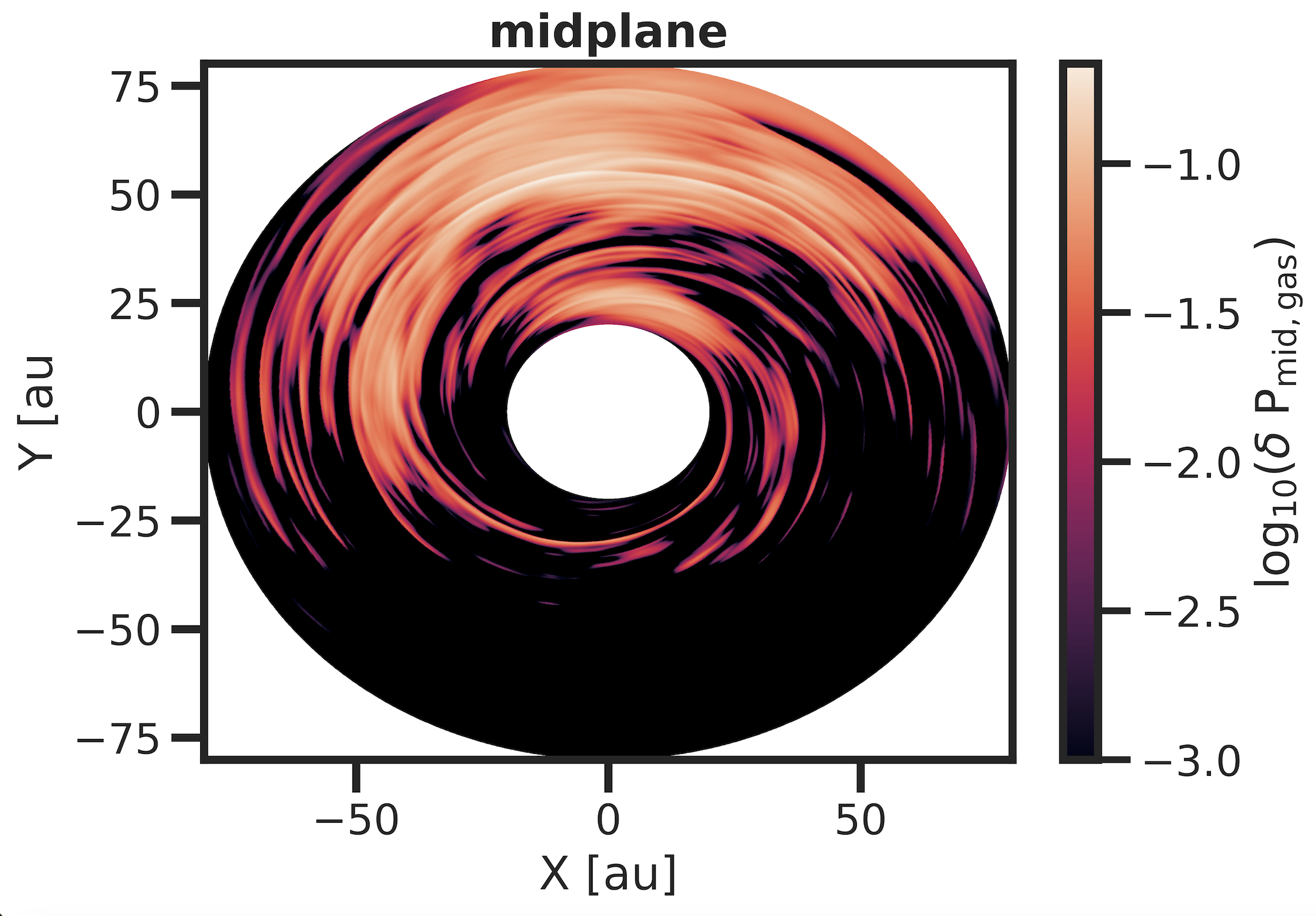}
\includegraphics[width=7.2cm]{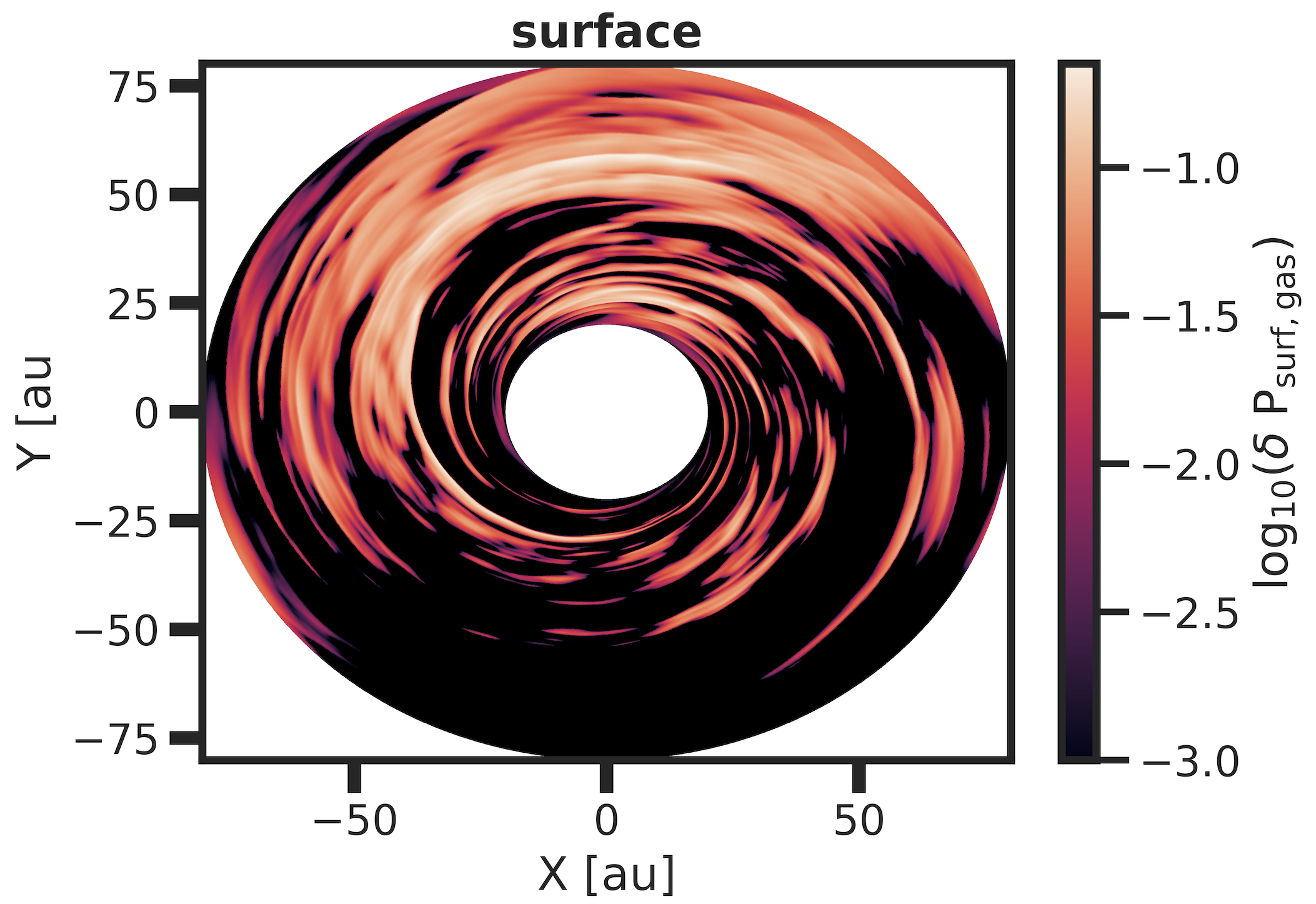}
\includegraphics[width=7.2cm]{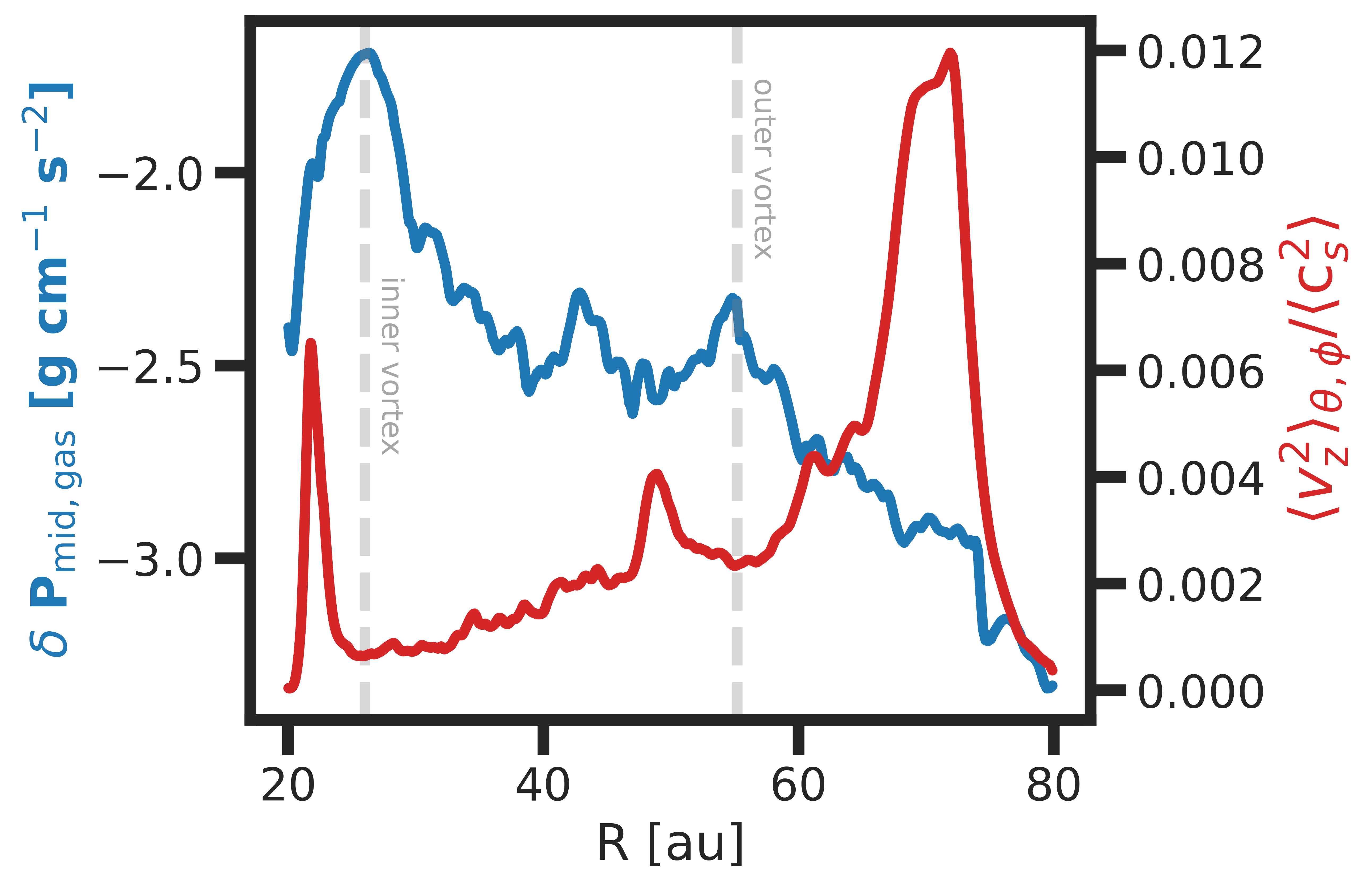}
\caption{Perturbed gas pressure structure in the midplane in logarithmic scale (top left) and in the surface (top right) of the global domain. Bottom panel shows the radial maximum gas pressure (blue curve) and the rms of the vertical velocity (red curve), $v_\mathrm{rms}$. At the location of the inner vortex and outer vortex we observe the pressure bumps following by a reduced $v_\mathrm{rms}$ demonstrating that vortices are weak turbulence features in the disk.}
\label{fig:pressure}
\end{figure*}

We examined the perturbed gas pressure in the disk, defined as $\delta P_\mathrm{gas} = \frac{P_\mathrm{gas}}{P_\mathrm{gas,0}}$, where $P_\mathrm{gas,0}$ is the initial pressure. As shown in the top panels of Figure \ref{fig:pressure}, the strongest gas pressure perturbations occur at the locations of the inner and outer vortices. The perturbed gas pressure map reveals an arc-like feature extending toward the lower right quadrant of the disk, seemingly originating from the outer vortex in the upper left, suggestive of a trailing spiral or filamentary structure driven by vortex-induced disturbances. The radial profile, shown in the bottom panel of Figure \ref{fig:pressure}, displays steep pressure gradients at the radial positions of both vortices, while the root-mean-square (rms) velocity, $v_\mathrm{rms}$ (plotted in red), exhibits local minima at the same locations. Similar steep pressure gradients were reported in \citet{Stadler_2025} for the  sources that exhibit dust crescents in the sub-mm continuum emission. 

\section{Dynamical trajectory of the particle}
\label{sec:3D_trajectory}

\begin{figure*}[htp!]
\centering
\includegraphics[width=7.5cm]{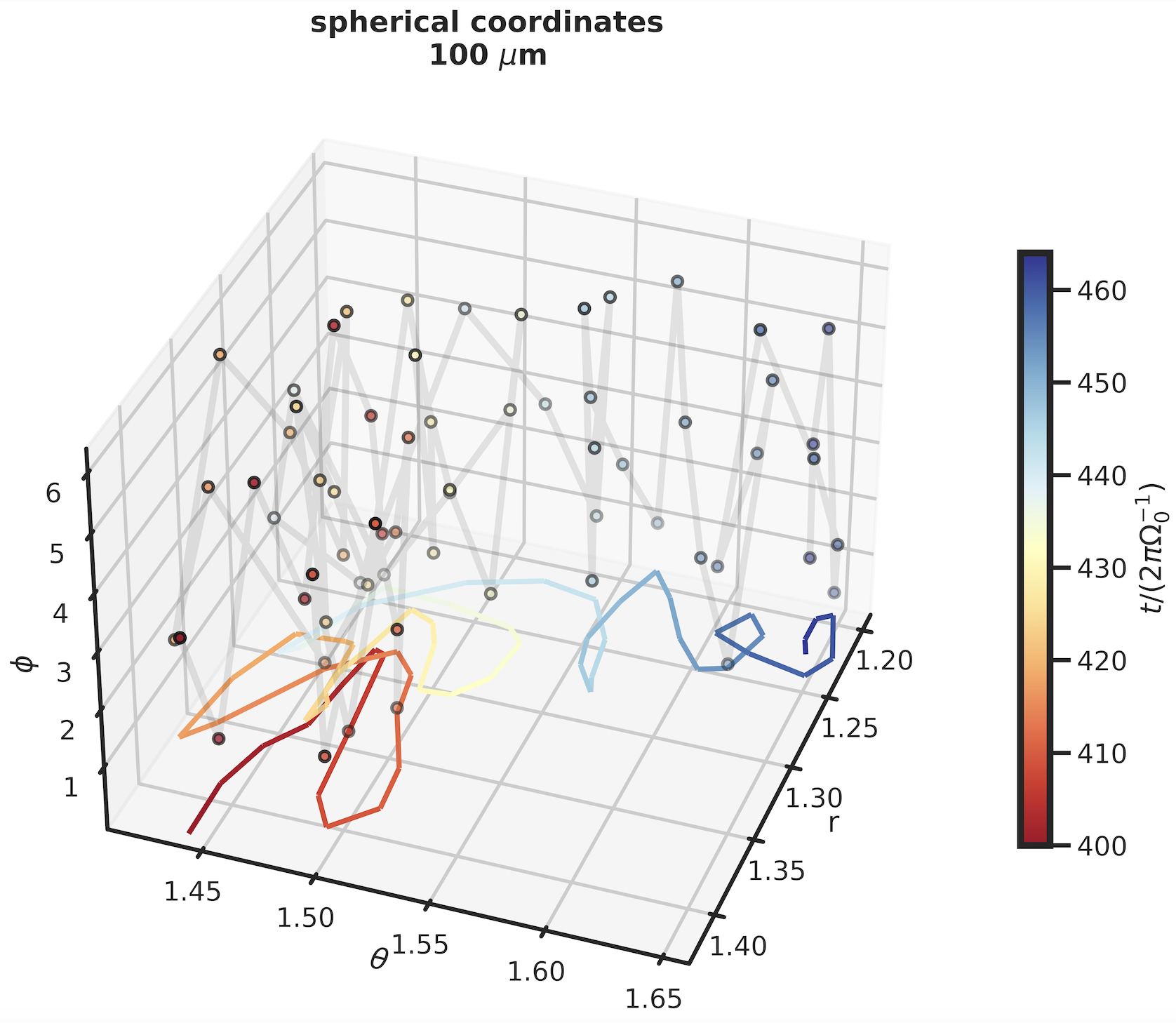}
\includegraphics[width=7.5cm]{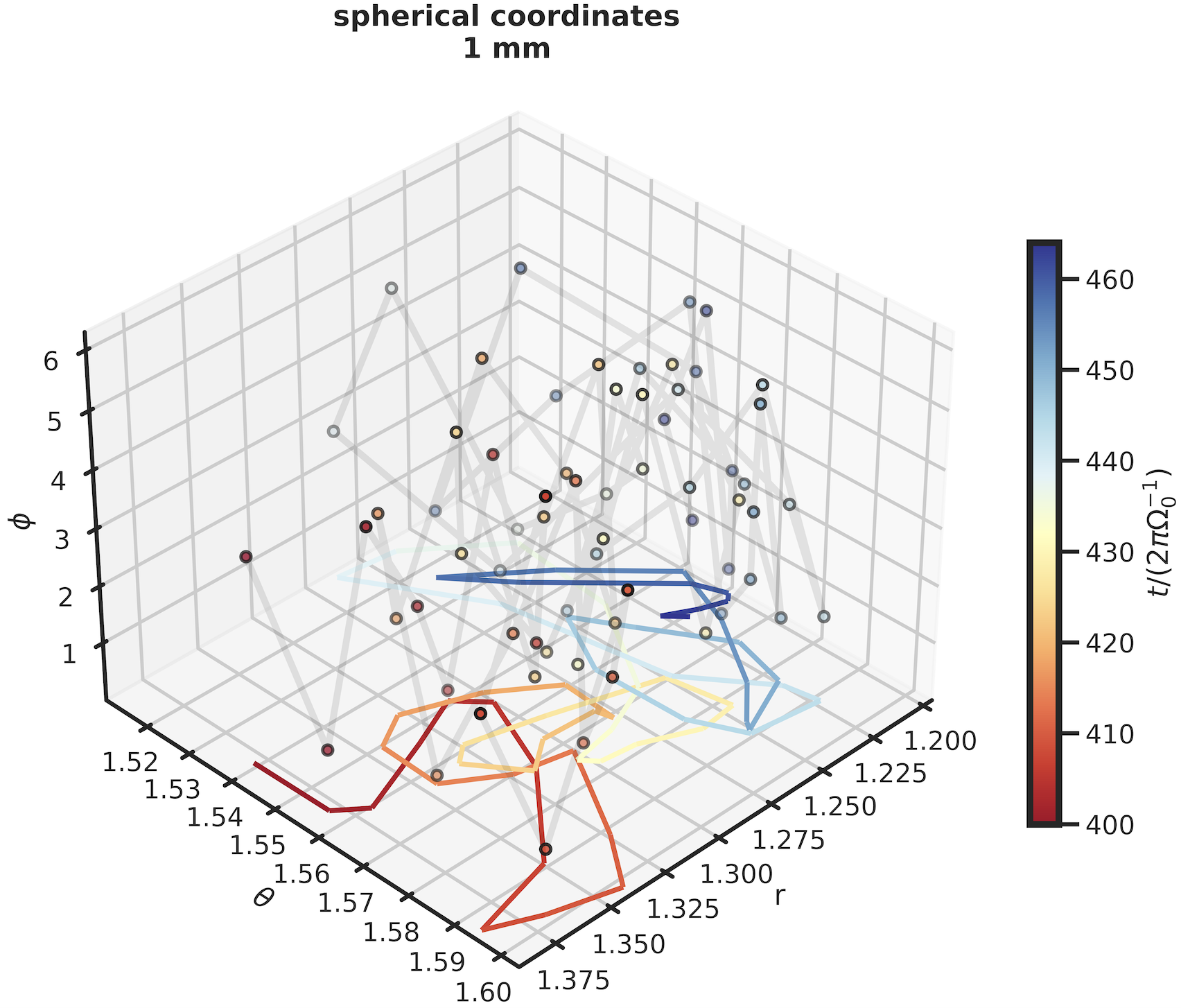}
\caption{Dynamical trajectory of a 1 mm-sized (left) and a 100 $\mu$m-sized (right) particle inside the outer vortex.}
\label{fig:part_trajectory}
\end{figure*}

We investigated the dynamical behavior of a particle once it is trapped by a vortex. A randomly selected particle, identified as a 1 mm-sized grain and a 100~$\mu$m-sized grain, was used to trace its 3D trajectory, as shown in Figure \ref{fig:part_trajectory}. The scatter markers, connected by light gray line segments, indicate the particle’s position at each orbit. The apparent jumps in the trajectory arise from the azimuthal angle wrapping from $\pi$ to $-\pi$ as the vortex completes full revolutions around the disk. In the $\theta$–$r$ plane, both particles follow looping paths within the vortex, while the vortex itself gradually migrates inward.

\end{appendix}

\end{document}